\documentclass[twocolumn,prd,superscriptaddress,amssymb,nofootinbib,amsmath,nobibnotes,aps,prlgroupedaddress]{revtex4-2}  
\usepackage{graphicx}  
\usepackage{dcolumn}   
\usepackage{bm}        
\usepackage{amssymb}   
\usepackage[]{amsmath,amssymb}
\usepackage{graphics,epsfig}
\usepackage{float}
\usepackage{mathtools}
\usepackage{esint}
\usepackage[utf8]{inputenc}
\usepackage[english]{babel}
\usepackage{mathtools}
\usepackage{xcolor}
\usepackage{float}  
\usepackage{setspace} 
\usepackage{indentfirst}  
\usepackage{cancel}  
\usepackage{soul}
\usepackage{hyperref}  
\hypersetup{
            colorlinks=true,
            linkcolor=black,
            urlcolor=blue,
            } 
\usepackage{tcolorbox}   

\usepackage{orcidlink}

\date{\today}

\begin{document}
\title{Cosmological perturbations with ultralight vector dark matter fields: numerical implementation in CLASS}

\author{Tomas Ferreira Chase \orcidlink{0009-0001-0286-2136}}
\email{tferreirachase@df.uba.ar}
\affiliation{Universidad de Buenos Aires, Facultad de Ciencias Exactas y Naturales, Departamento de Física. Buenos Aires, Argentina.} 
\affiliation{CONICET - Universidad de Buenos Aires, Instituto de Física de Buenos Aires (IFIBA). Buenos Aires, Argentina}

\author{Matías Leizerovich \orcidlink{0000-0002-6438-2285}}
\email{mleize@df.uba.ar}
\affiliation{Universidad de Buenos Aires, Facultad de Ciencias Exactas y Naturales, Departamento de Física. Buenos Aires, Argentina.} 
\affiliation{CONICET - Universidad de Buenos Aires, Instituto de Física de Buenos Aires (IFIBA). Buenos Aires, Argentina}

\author{Diana Lopez Nacir \orcidlink{0000-0003-4398-1147}}
\affiliation{Universidad de Buenos Aires, Facultad de Ciencias Exactas y Naturales, Departamento de Física. Buenos Aires, Argentina.} 
\affiliation{CONICET - Universidad de Buenos Aires, Instituto de Física de Buenos Aires (IFIBA). Buenos Aires, Argentina}

\author{Susana Landau \orcidlink{0000-0003-2645-9197}}
\affiliation{Universidad de Buenos Aires, Facultad de Ciencias Exactas y Naturales, Departamento de Física. Buenos Aires, Argentina.} 
\affiliation{CONICET - Universidad de Buenos Aires, Instituto de Física de Buenos Aires (IFIBA). Buenos Aires, Argentina}

\begin{abstract}
In this work we consider a dark matter candidate  described by an ultralight vector field,  whose mass is  in principle in  the range   $H_{\rm{eq}}\sim 10^{-28}\rm{eV}\ll m< \rm{eV}$.  The homogeneous  background vector field is assumed to point in a given direction.
We present a numerical implementation of cosmological perturbations in a Bianchi type I geometry with vector field dark matter in a modified version of the Cosmic Linear Anisotropy Solving System (CLASS).  We study the  evolution of large-scale cosmological perturbations in the linear regime. We  compute the  matter power spectrums defined for Fourier modes pointing in a given direction. We obtain interesting features in the  power spectrums  whose observational significance depends   on the field mass. We compare the results with the standard $\rm{\Lambda CDM}$ and with the corresponding well-studied ultralight scalar field dark matter case.  
As for the scalar case we obtain a suppression in the  power spectrums  at small scales characterized by the same scale, namely the Jeans scale. The main characteristic feature of the vector field model we notice here for first time is  that the amplitude of the suppression effect depends on the direction of the Fourier modes with respect to the background vector field, leaving eventually a possible anisotropic imprint in structure formation at small scales.
\end{abstract}
\maketitle

\section{Introduction}

A great amount of data supports the existence of dark matter  (DM) as a key component of our universe. The standard cosmological model, 
$\Lambda$\rm{CDM}, treats dark matter as a distribution of non-relativistic particles   that (once produced in the early universe) evolve practically without interacting with other constituents, apart from the coupling through gravity.  The parameter  of the model that determines the abundance of CDM  is  $\Omega_{\rm{CDM}}$, which is defined as the  ratio of the CDM energy to the total energy in the universe.   From Cosmic Microwave Background (CMB) observations the value $\Omega_{\rm CDM}= 0.265 \pm 0.005$ is inferred \cite{Planck:2018vyg}.
The  CDM model   allows for an explanation of a large number of astrophysical and cosmological phenomena \cite{Weinberg:2008zzc,Dodelson:2020bqr}.  So far only the large-scale gravitational interaction of the dark matter has been detected and there are a large number of alternative models that could also be viable.

Among the most studied alternative DM models in recent literature are those in which  DM is described by an ultralight (with a mass smaller than $\sim {\rm eV}$) field (ULDM).  The most popular ULDM candidates are scalar fields (spin 0), mainly  ultralight axions (ULAs) (see for example \cite{Preskill:1982cy,Turner83, Marsh:2015xka, Hui:2016ltb,OHare:2024nmr}). However, vector (spin 1) and tensor (spin 2) ULDM models could be in principle as  viable as the scalar ones and are  also being   considered in the literature (see for instance \cite{Nelson:2011sf, Arias:2012az,Cembranos:2016ugq,Marzola:2017lbt}).  An important question is whether or not it is possible to distinguish between these models. It is clear that to investigate this it is necessary to identify the relevant observable properties and  carry out a calculation of the corresponding predictions,  on all scales where the models make calculable predictions that can be tested observationally or experimentally.    

It is known that there are relevant observational and experimental data involving cosmological, astrophysical and laboratory scales, which can be used to discriminate between alternative models. For ULDM models with a given spin and in certain mass-ranges,    predictions have been studied on several scales and  some of them have been used  to probe and constrain  the candidate models  
 (see for instance   \cite{Khmelnitsky:2013lxt, Nomura:2019cvc, Armaleo:2020yml,Blas:2016ddr,Blas:2019hxz,LopezNacir:2018epg,
 Armaleo:2019gil,Stott:2020gjj, Hlozek:2014lca, Urena-Lopez:2015gur, Hlozek:2017zzf, Lague:2021frh,Rogers:2020ltq, Ferreira:2020fam}  and  references  therein). The next generation of experiments (such as EUCLID\footnote{ http://sci.esa.int/euclid/}, LSST\footnote{https://www.lsst.org} or SKA\footnote {https://skatelescope.org}) will provide an unprecedented quantity and quality of observational data; for example, those corresponding to observables that characterize the statistical properties of the distribution of galaxies, neutral hydrogen and the effect of gravitational lensing. These observables are sensitive to the details of the structure formation process and offer a window to test candidate models. 

 With regard to the large-scale cosmological perturbations, while scalar ULDM (SFDM) models have been broadly studied and tested using cosmological data \cite{Hlozek:2014lca, Urena-Lopez:2015gur, Hlozek:2017zzf, Lague:2021frh,Rogers:2020ltq},  the study of cosmological perturbations in vector ULDM  models  remain on theoretical grounds \cite{Cembranos:2016ugq,Chase:2023puj}.

 With the goal of moving forward  in this direction in this paper, as done in  \cite{Chase:2023puj}, we consider models where the inferred DM abundance is explained by the presence of a homogeneous ULDM vector field (with the appropriate amplitude) pointing in a given direction, which only interacts gravitationally with the standard model particles. We refer to  these models as VFDM models. Such vector field could be produced during inflation (see for instance \cite{Arias:2012az, Nakayama:2019rhg,Kaneta:2023lki,Kitajima:2023fun}).

 We focus on studying the predictions of the models with respect to the evolution of large-scale cosmological perturbations. 
In this VFDM scenario,  the background solution of Einstein equations corresponds to a Bianchi type I universe where the anisotropies are described by the   \textit{shear tensor} of the metric, $\sigma_{ij}$, generated by the  VFDM  \cite{Chase:2023puj}.  
Even though the VFDM background breaks isotropy,
when the characteristic timescale of the evolution of the background metric (given by the inverse of the Hubble rate $H$) is greater or equal  to  the inverse of the boson mass, the field  oscillates, and for not too light fields (we will start quantifying this below) the background anisotropies generated by the VFDM may not leave a significant imprint on cosmological observables. This is in fact expected from  general isotropy theorems \cite{Cembranos:2012kk}. In this way there could be a dynamical approaching to an effectively isotropic metric,    making  the  VFDM model a viable dark matter candidate for a given mass range. Hence, without taking any other observational constraint (since the Hubble rate decreases with time in radiation and matter domination eras), the possible masses are such $H_{\rm{eq}}\sim 10^{-28}\rm{eV}\ll m$, where $H_{\rm{eq}}$ is the value of $H$ at \textit{equality}, when the abundance of the DM component (here the VFDM) equals the abundance of radiation.
 
 Beyond cosmological observational probes,  complementary constraints on the possible masses (in a window around $10^{-23}$ eV) of  VFDM interacting only throughout gravity could  be obtained with the use of pulsar timing data  \cite{Khmelnitsky:2013lxt,Nomura:2019cvc}. 
For larger values of the mass of the vector field, regardless of whether its abundance accounts for the whole DM or not,    constraints can be obtained from the study of astrophysical black holes, because bosonic condensates can form  in the surrounding of  spinning black holes via superradiant instabilities  \cite{Baryakhtar:2017ngi,Pani:2012vp}.
Such studies lead to the following exclusion windows for the mass of the vector field  at 68\% confidence limit \cite{Stott:2020gjj}: {$6.2\times10^{-15}  {\rm eV} \leq m \leq 3.9\times 10^{-11}  {\rm eV}, \, 2.8\times 10^{-22} {\rm eV} \leq m \leq 1.9\times10^{-16} {\rm eV}$}, which incorporate   observations such as the event GW190521 and the shadow of M87*.
 
 In \cite{Chase:2023puj} 
 some of us  derived all the  equations needed to evolve the scalar sector of the cosmological perturbations in VFDM models  in the linear regime (both in   synchronous gauge and in   Newtonian gauge), neglecting the vector and tensor modes, which are defined according to the  standard scalar-vector-tensor (SVT) decomposition for cosmological perturbations \cite{Weinberg:2008zzc}. 
 In \cite{Chase:2023puj}  a derivation of the initial conditions corresponding to the  so-called adiabatic mode is also provided.   Assuming adiabatic initial conditions, in the same paper it was shown  that the metric shear tensor must be taken into account in Einstein's equations in the early universe, at least for a range of masses ($m < 10^{-22} {\rm eV}$).
Otherwise, there would be a large infrared contribution from VFDM perturbations at the linear level  at early times, when the radiation is supposed to be dominant, that would be misinterpreted as producing an infrared effect on cosmological observables.  
Indeed, it was shown that the contribution of the shear tensor cancels that corresponding to the VFDM perturbations in the infrared limit.

 In this paper we present a numerical implementation in CLASS code \cite{Blas:2011rf,Lesgourgues:2011rh} of the system of equations  presented in \cite{Chase:2023puj} and we study the evolution of the perturbations assuming adiabatic initial conditions.

The  paper is organized as follows. In section \ref{sec:Vector field background} we present the VFDM model and study  the  background evolution. Besides, we discuss the implementation in CLASS of the background equations and the Bianchi I metric. 
In section \ref{sec:Vector field perturbations} we focus on the cosmological perturbations  at linear order with VFDM, and describe the procedure used while implementing the equations for the scalar perturbations in CLASS including the adiabatic initial conditions derived in \cite{Chase:2023puj}.   We call the modified version of  CLASS v3.2 as \textit{class.VFDM}, which we make public\footnote{\url{https://github.com/classULDM/class.VFDM}}. In Sec. \ref{sec:Cosmological observables} we study the evolution of the VFDM perturbations and the resulting anisotropic  matter power spectrum for different directions of the Fourier modes with respect to the direction of the background vector field. We discuss on the results obtained with the code by   using  several analytic approximations. Finally, in section \ref{sec: conclusions} we summarize the conclusions of our work. Appendices \ref{appendix_fluid_variables} and \ref{appendix:change_of_variables} contain additional intermediate equations and details of the calculations. In Appendix \ref{sec:shear_impact}  we present an analysis illustrating the importance of the metric shear $\sigma_{ij}$ on the observables. In Appendix \ref{sec:jeans_scale} we provide an alternative derivation of the Jeans mechanism in VFDM.

\section{ VFDM model and Background evolution} \label{sec:Vector field background}

We consider the dynamics of  dark matter is described by the following action for an ultralight vector field $A^{\mu}(\tau,\vec{x})$ in General Relativity,
\begin{equation}
    S = -\int d\tau\,dx^3\,\sqrt{-g}\left[\frac{1}{4} F^{\mu\nu}F_{\mu\nu} + \frac{m^2}{2} A^{\mu}A_{\mu}\right]\, ,
\end{equation}
where $F_{\mu\nu} = \nabla_{\mu}A_{\nu} - \nabla_{\nu}A_{\mu}$ is the usual field tensor, $g$ is the determinant of the metric and $m$ the vector's mass. The corresponding equation of motion for the vector field can be obtained varying the action with respect to the field, thus obtaining the Proca equation
\begin{equation}
    \nabla_{\nu} F^{\mu \nu} + m^2 A^{\mu} = 0\, .
    \label{general_eq_motion}
\end{equation}

In this work we study VFDM to linear order, so we write the field as a combination of a background field which is homogeneous but not isotropic, and a perturbation $A^{\mu}(\tau,\vec{x}) \rightarrow A^{\mu}(\tau) + \delta A^{\mu}(\tau,\vec{x})$. In this section we focus on the background quantities. We assume that the background vector field is pointing in a given direction, so we can parameterize it as $\vec{A} = A(\tau) \hat{A}$. 

Vector field dark matter requires an anisotropic background metric for the model to be consistent \cite{Chase:2023puj}. 
We then generalize the FLRW (Friedmann-Lema\^itre-Robertson-Walker) metric to a Bianchi I metric to model these anisotropies. The metric is given by
\begin{equation}
    ds^2 = a(\tau)\left[-d\tau^2 + \gamma_{ij} dx^i dx^j\right]\,,
\end{equation}
where $\tau$ is the so-called  conformal time (which is related to the cosmic time $t$ by $dt= a(\tau) d \tau$),  $a(\tau)$  the scale factor, $\gamma_{ij} = e^{-2\beta_i(\tau)} \delta_{ij}$ and the functions  $\beta_i$ (with $i=1,2,3$) are constrained by $\sum_i^3 \beta_i = 0$. The relevant quantity describing the anisotropies is the shear tensor, defined as $\sigma_{ij} = \frac{1}{2}\left(\gamma_{ij}\right)^{.}$ \cite{Pereira:2007yy}, where dot derivatives are with respect to conformal time.

The vector equations (\ref{general_eq_motion}) can be split in a dynamical equation for the spatial components $\vec{A}$, and a constraint equation for the time component $A^{0}$. We suppose that the background vector is pointing in a given direction. Due to the anisotropic nature of the background metric, the time derivative of the vector direction $\hat{A}$ does not vanish but is of order $\mathcal{O}(\sigma_A)$. To leading order in $\sigma_A$ we neglect the time derivatives of $\hat{A}$, and focus on an equation of motion for the vector modulus $A(\tau)$. Then, the equations of motion for the background vector field are \cite{Chase:2023puj}
\begin{equation}
    A_0 = 0 \,, \quad\quad \ddot{A} + m^2 a^2 A = 0 \, .
    \label{eq_mov_A_0}
\end{equation}

An approximate solution of these equations can be found  in two regimes determined by the VFDM mass. When $m a \ll \mathcal{H}$ (which is equivalent to $m  \ll H$ using the Hubble rate in cosmic time) the equation of motion can be approximately solved by a growing mode proportional to the scale factor, while for $ ma \gg \mathcal{H}$ the equation can be solved under a WKB (Wentzel, Kramers, Brillouin) approximation,
\begin{equation}
    A(\tau) \propto 
    \begin{cases}
        a \qquad\qquad\qquad\qquad\,\,\,\,\,\, \quad a < a_{osc}\\
        a^{-\frac{1}{2}} \cos\left(\int m a \, d\tau\right) \,\quad\,  a_{osc} < a 
    \end{cases},
    \label{background_field_evolution}
\end{equation} 
\noindent where $a_{osc}$ is defined by $m a=m a_{osc}=\mathcal{H}$ and corresponds to the time when the field starts oscillating.

With this solutions we can calculate the fluid variables, that is the energy density $\rho_A$, pressure $P_A$ and shear $\Sigma^i_j$, defined from the stress tensor of the field (see Appendix \ref{appendix_fluid_variables}). We treat the shear tensor perturbatively. We work at zero order in the shear in the fluid variables, and to linear order on the left-hand side (l.h.s) of Einstein equations. For the energy density of the field we have that
\begin{equation}
    \rho_A = \frac{\dot{A}_{i}^2+ m^2a^2 A_{i}^2}{2 a^2} \propto \begin{cases}
    a^{-4} \quad m a\ll \mathcal{H} \\
    a^{-3} \quad m a \gg\mathcal{H}
    \end{cases},
    \label{eq_background_energy_density}
\end{equation}
while for the pressure $P_A$ we get a radiation-like equation of state before the field starts oscillating ($w_A = \frac{1}{3}$ for $m a\ll \mathcal{H}$) and a CDM equation of state after the field starts oscillating ($w_A = 0$ for $m a \gg \mathcal{H}$). The fact that $\rho \propto a^{-4}$ when $m a \ll \mathcal{H}$ is an important difference with respect to SFDM  models, where the density is constant until the field starts oscillating. Finally, with the field solution we can calculate the background shear of the vector field,
\begin{equation}
    {\Sigma^i}_j =  
        - 6 P_A \left(\hat{A}_i \hat{A}_j - \frac{\gamma_{ij}}{3} \right),  
\end{equation} whose evolution is determined by the one of the pressure. 

To complete the set of equations for the background we have to consider Einstein equations. For a Bianchi I geometry these equations give the generalized Friedman equation, which  involves  a term containing the shear tensor,
\begin{equation}
    \mathcal{H}^2 = \frac{\sigma^2}{6} +  \frac{a^2}{3m_P^2} \rho_T\,,
    \label{modified_friedmann}
\end{equation}
and the equation for the shear tensor (from the spatial traceless part) sourced by the vector shear\footnote{The vector field is the only species we consider here that has anisotropies at the background level.},
\begin{equation}
    ({\sigma^i}_j)^{\dot{}} + 2 \mathcal{H}\, {\sigma^i}_j = \frac{a^2}{m_P^2} {\Sigma^i}_j\,,
    \label{metric_shear_eq}
\end{equation}
where $m_P$ is the Planck mass.

In analogy to the definition of the abundance of any standard specie, $\Omega_{i} =\rho_{i}/\rho_c$ (with index $i$ runing   over every species other than the vector field), where $\rho_c= 3 m_P^2 \mathcal{H}^2/a^2$, 
we define the VFDM abundance, 
${\Omega}_A=\rho_{A}/\rho_c$, and also  the \textit{shear abundance} as 
\begin{equation}
    \Omega_{\sigma} = \frac{\sigma^2}{6\mathcal{H}^2}  \,.
    \label{eq_shear_abundance_definition}
\end{equation} 
Assumimg there is no significant initial  anisotropies before the initial time $a_{\rm ini}$, we can solve Eq. (\ref{metric_shear_eq}) perturbatively in the anisotropies, and calculate the shear abundance generated by the vector field in the different eras, for $a\gg a_{\rm ini}$. Then, since at leading order we can neglect the time derivatives of the versors,  the tensorial structure of the shear is the same as the one of the source,
\begin{equation}
    {\sigma^i}_j =  
         \frac{3}{2} \sigma_A \left(\hat{A}_i \hat{A}_j - \frac{\gamma_{ij}}{3} \right),  
\end{equation}
where we defined $\sigma_A=\hat{A}_i\hat{A}_j\sigma^{ij}$. Notice $\sigma^2=3 \sigma_A^2/2$. In what follows we   use $\sigma_{A}$ as one of the dynamical variables. 
Before the field starts oscillating the shear abundance is approximately constant. 
This is a difference with respect to Bianchi I models without sources on the right-hand side (r.h.s) of Eq.(\ref{metric_shear_eq}), where it can be shown the shear abundance decays as $a^{-2}$ in radiation era. In this case, the shear abundance starts decaying only after $a>a_{osc}$ (when the source in Eq.(\ref{metric_shear_eq}) averages to zero), behaving as
\begin{equation}
    \Omega_{\sigma} \propto
    \begin{cases} 
  \,\, a^{-2} \,\quad\qquad a_{osc} < a < a_{eq}   \,  \\[4pt]
  \,\, a^{-3} \,\,\quad\qquad a_{eq} < a \,
    \end{cases}  .
    \label{eq_shear_abundance_evolution}
\end{equation} 

In the implementation of the model in CLASS we use  the vector abundance as a dynamical variable. Then, we write the continuity equations of the background Einstein equations as 
\begin{align}
    \dot{\mathcal{H}} &= -\frac{1}{2} (1+3w_T) \mathcal{H}^2 (1-\Omega_{\sigma})-2  \mathcal{H}^2 \Omega_\sigma \,,     \label{friedman_1}\\[4pt]
    \dot{\Omega}_A &= 3\mathcal{H}(w_T - w_{A}) \Omega_{A} 
    \,, \label{abundancia_1}
\end{align}
where $\rho_T$ is the total energy density and $w_T$ is the total equation of state, namely
\begin{equation}
    w_T = \frac{P_T}{\rho_T} = \frac{1}{\rho_T} \sum_{I= i,A} w_I \,\rho_I \,.
\end{equation}

The equations of motion of the vector field (Eq. \ref{eq_mov_A_0}), the modified Friedmann equation (Eq. \ref{modified_friedmann}), the spatial-traceless Einstein equation (Eq. \ref{metric_shear_eq}) and the continuity equations (Eqs. \ref{friedman_1} and \ref{abundancia_1}) conform the set of background equations of the model that were implemented in CLASS, as   described next.

\subsection{Implementation in CLASS} \label{sec:implementation_in_class}

In order to solve the dynamical background equations  numerically we follow the procedure developed in \cite{Urena-Lopez:2015gur} for SFDM models. First, we transform Eq. (\ref{eq_mov_A_0}) into a system of two differential equations of first order with the following change of variables,
\begin{equation}
    \begin{cases}
    \vec{A} = \sqrt{6}\, m_P \frac{\mathcal{H}}{m} \sqrt{\Omega_A} \sin \left(\frac{\theta}{2}\right) \hat{A} \\[4pt]
    \dot{\vec{A}} = \sqrt{6}\, m_P \, a\mathcal{H} \sqrt{\Omega_A} \cos \left(\frac{\theta}{2}\right) \hat{A}
    \end{cases},
\end{equation}
where $\theta$ is a new variable.

We can see that in terms of these new variables, the equation of state reads
\begin{equation}
    w_A = \frac{1}{3} \cos(\theta)\,,
\end{equation}
and the system of equations for the background dynamics reduces to
\begin{equation}
    \begin{cases}
    \theta' = \sin(\theta)+y \\[4pt]
    \Omega_{A}' = 3(w_T - w_A) \Omega_{A}  \\[4pt]
    y' = \frac{3}{2}(1+w_T)y \\[4pt]
    \sigma_A^{\prime} = -2 \sigma_A - 12 w_A \mathcal{H} \Omega_A 
    \end{cases},
    \label{background_system_of_eqs}
\end{equation}
where the prime derivatives means  $\frac{d}{d \log a}$ and $y =2  {m a}/{\mathcal{H}}= 2 {m }/{H}$. The previous system of equations is the one implemented for the background in our modified version of CLASS.

Now we study the system at different time scales. This is useful for gaining insight in what to expect from the numerical solutions, and also to calculate the initial conditions of the system. We consider two stages of the evolution. First, we have an early time epoch, where we assume a radiation dominated era. Then, we study the system at late times when the field is highly oscillating, and we assume a matter dominated era. Finally, in Sec. \ref{sec:threshold_discussion} we present some general considerations of the implementation in CLASS of the model. We use the subindex $ini$ to denotes the quantity evaluated at the initial time of integration, and also use the subindex $0$ to denote the quantity evaluated at present time.

\subsubsection{Initial conditions} \label{sec:background_initial_conditions}

At early times we have that $H_{\rm ini} \gg m$ for the masses considered in this work ($H_{\rm ini} \gg m > H_{\rm eq}$). From the field solution at early times (Eq. \ref{background_field_evolution}) we see that $A^{\prime} = \mathcal{H} A$, so $A^{\prime} \gg m a A$. In terms of the new variables, this means $\theta \ll 1$. At early times we also assume a radiation era, so $w_T \sim \frac{1}{3}$. Then, the system of equations reads
\begin{equation}
    \begin{cases}
    \theta' = \theta + y \\
    \Omega_{A}' = 0 \\
    y' = 2 y
    \end{cases}.
\end{equation}

We can easily solve for $y$ and $\Omega_A$, and then for $\theta$. Then, we obtain
\begin{equation}
    \begin{cases}
    \theta = \dfrac{2m}{H_{\rm ini}}\left(\dfrac{a}{a_{\rm ini}}\right)^2 \\[4pt]
    \Omega_{A} = const \equiv \Omega_{A,{\rm ini}} \\[4pt]
    y = \theta  
    \end{cases}.
    \label{early_time_solution}
\end{equation}

Assuming that at early times the radiation component is the dominant one  (for this analytical estimate we neglect   the shear abundance), by using the modified Friedman equation (Eq. \ref{modified_friedmann}) at early times $H_{\rm ini} = H_0 \Omega_{r,0}^{1/2}/a^2$,  where $\Omega_{r,0}$ is the radiation abundance at present time, we can write the initial condition for $\theta$ as
\begin{equation}
    \theta_{\rm ini} = \frac{2 m\, a_{\rm ini}^2}{H_0 \Omega_{r,0}^{1/2}}\,.
\end{equation}

For the masses considered in this work the field starts oscillating in the radiation era. Then, to calculate the initial vector abundance we use that $\Omega_{A,\rm ini} \sim \Omega_{A,\rm{osc}}$, where $a_{\rm{osc}}$ is such that $H(a_{\rm{osc}}) = m$. For $a>a_{\rm{osc}}$ we have that  
\begin{equation}
    \rho_{A} = {\rho_{A}}_{osc} \left(\frac{a_{osc}}{a}\right)^3  \,,\qquad \rho_{r} = {\rho_{r}}_{osc} \left(\frac{a_{osc}}{a}\right)^4 \,.
\end{equation}

Then, by evaluating at present time we obtain
\begin{equation}
    \Omega_{A,{\rm ini}} = \frac{\Omega_{A,0}}{\Omega_{r,0}} a_{osc}\,,
    \label{omega_initial_condition}
\end{equation}
where in the last step we used that $\Omega_{r,osc} \sim 1$ and 
\begin{equation}
    a_{osc} \sim \left( \frac{\Omega_{r,0}^{1/2} H_0}{m} \right)^{1/2}\,.
\end{equation}

\subsubsection{Late times}

Now we solve Eqs. (\ref{background_system_of_eqs}) at late times in the matter dominated era. If we assume a cold dark matter equation of state $w_T \sim 0$, we can solve for $y$ at late times as 
\begin{equation}
    y = y_{eq} \left(\frac{a}{a_{eq}}\right)^{3/2}
\end{equation}
where $eq$ denotes the quantity evaluated at equality. 

Now we can solve for $\theta$. After the field starts oscillating we have that $y = 2 m/H \gg 1$. Then, as $\sin(\theta) \leq 1$, we have that $ \theta' \sim y$. Solving for $\theta$ we obtain
\begin{equation}
    \theta = \frac{2}{3} y \gg 1.
\end{equation}

We then have that the field is highly oscillating in the matter era, as in the WKB approximation of Eq. \ref{background_field_evolution}. In the matter era we then average the equations on cosmological time scales. 
For example, we can see that for late times the vector field behaves as CDM on this scales, since 
\begin{equation}
   \langle w_{A} \rangle \sim \frac{1}{3}\langle \cos\theta \rangle = 0\,,
\end{equation}
where the average $\langle \dots \rangle$ is taken over cosmological timescales and in the limit $m a\gg  \mathcal{H}$. In Sec. \ref{sec:threshold_discussion} we discuss an implementation of this time-average in the code.

\subsubsection{Average procedure and shooting algorithm} \label{sec:threshold_discussion}

The equation of motion for the VFDM (Eq. \ref{eq_mov_A_0} or Eq. \ref{background_field_evolution}) suffers from stiffness problems when $m \gg H$, so numerical integrator programs are likely to be unstable. To fix this, we average the equations after the field starts oscillating by implementing an analytic cut-off\footnote{For a discussion about this cut-off procedure see Appendix A in \cite{Urena-Lopez:2015gur}.} in the trigonometric functions,
\begin{equation}
    \begin{cases}
    \sin(\theta) \longrightarrow \frac{1}{2}\left(1-\tanh(\theta^2-\theta_0^2)\right) \sin(\theta) \\
    \cos(\theta) \longrightarrow  \frac{1}{2}\left(1-\tanh(\theta^2-\theta_0^2)\right) \cos(\theta)
    \end{cases}.
    \label{stiff_trig}
\end{equation}

Then, when $\theta > \theta_0$ the previous functions rapidly approach to 0. We use a threshold of $\theta_0 \sim \mathcal{O}(100)$.
Eq. (\ref{omega_initial_condition}) is not the correct initial condition for the abundance precisely. In order to obtain the desired VFDM abundance at present time, we finely tuned $\Omega_{\rm ini}$ with the shooting algorithm already present in CLASS for scalar models. That is, we wrote $\Omega_{\rm ini}\to A\cdot \Omega_{\rm ini}$, where $A$ is a constant of order $\mathcal{O}(1)$ determined by the shooting algorithm. The shooting algorithm picks the constant $A$ such that the vector abundance at present time is the desired one.

One final remark is that the code also needs to compute the energy density of the field in every step of the evolution to source the Friedmann equation. Since we are using the VFDM abundance as a dynamical variable, we write the energy density as
\begin{equation}
    \rho_A = \frac{\Omega_A}{1 - \Omega_A} \left(\rho_r + \rho_b + \rho_{\Lambda}\right)\,,
\end{equation}
where $b$ stands for the barions and $\Lambda$ for the cosmological constant.

\subsection{Results}\label{sec:background_results}

\begin{figure}[th]
    \centering 
    \includegraphics[width=0.47\textwidth]{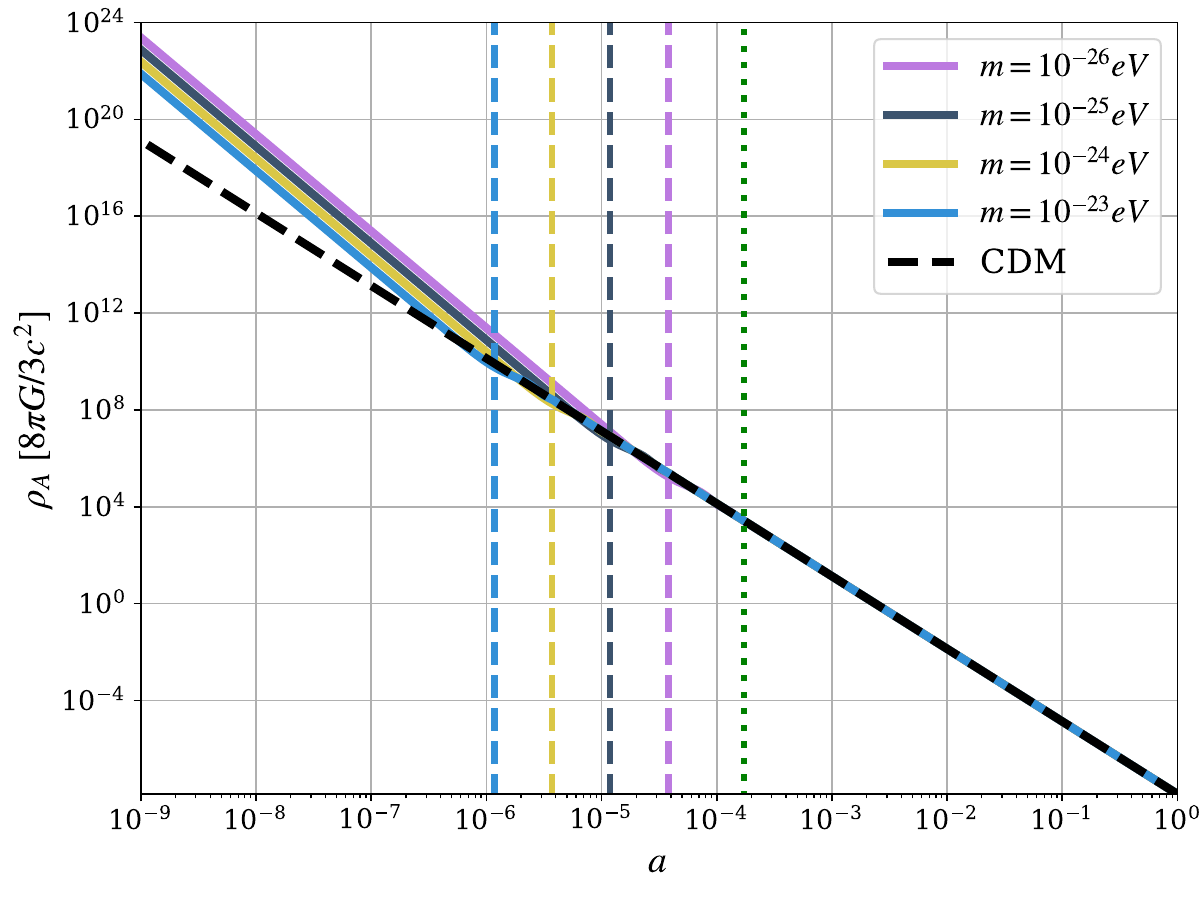}
    \includegraphics[width=0.47\textwidth]{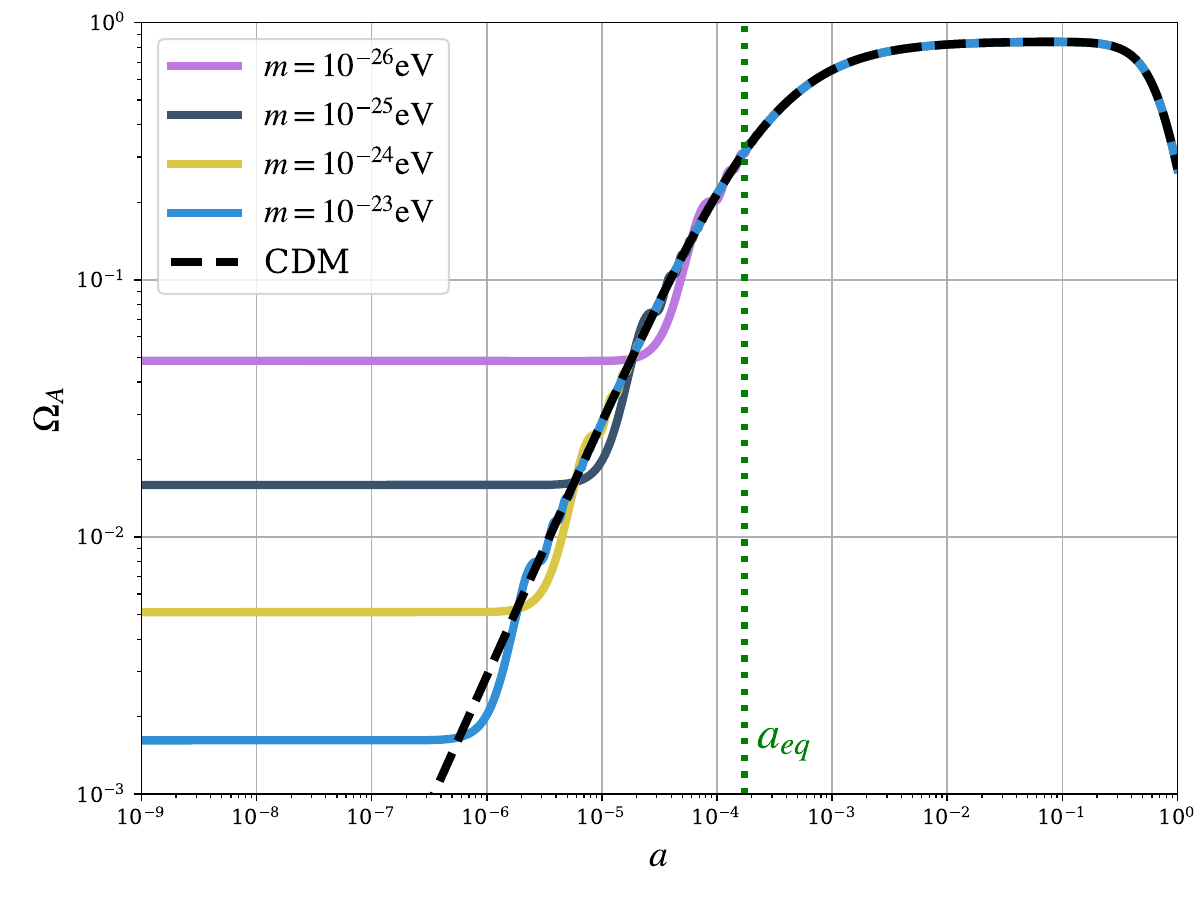}
    \caption{Numerical solutions for the VFDM for different masses. On top, we have the VFDM energy density for different masses as a function of the scale factor. The dashed vertical lines show  the time $a_{osc}$ when the field starts oscillating, where $H(a_{osc})=m$. On the bottom, we have the VFDM abundance. We can see that the VFDM follows the CDM behaviour once the field starts oscillating.}
    \label{fig:vector_background}
\end{figure}

In Fig. \ref{fig:vector_background} we can see the VFDM energy density (top) and abundance (bottom) for different masses of the field calculated with CLASS. In the top panel we can see that the vector's energy density has two different behaviours depending on whether  the field mass is smaller than  $H$ or larger, as expected from Eq. (\ref{eq_background_energy_density}). The value of $a=a_{osc}$ for which $H(a_{osc})=m$ are indicated as vertical dashed lines. We can see that once the field starts oscillating it behaves as cold dark matter.

\begin{figure}[th]
    \includegraphics[width=0.47\textwidth]{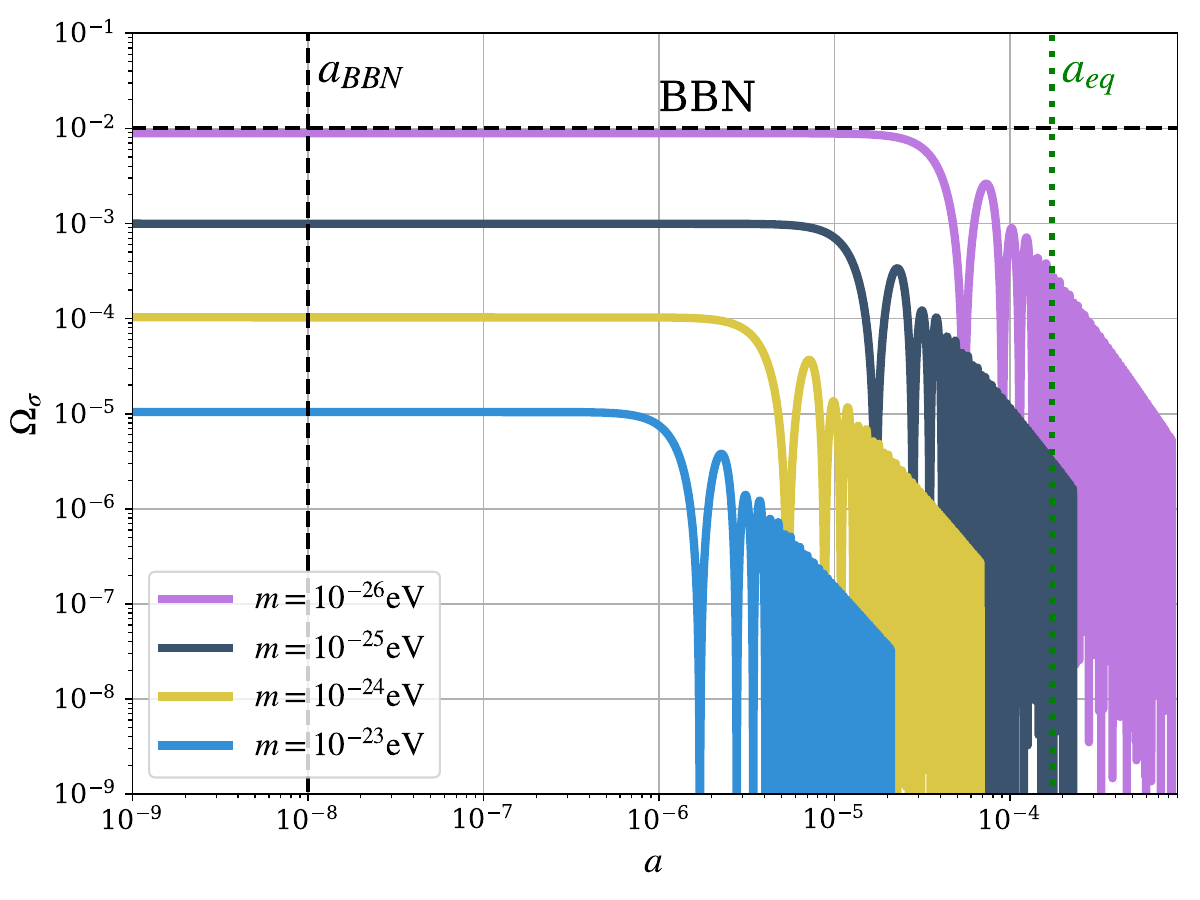}
    \caption{Numerical solutions for the shear abundance defined as in Eq. (\ref{eq_shear_abundance_definition}), $\Omega_{\sigma} = 3\sigma_A^2/2\mathcal{H}^2$. The shear was calculated perturbatively with Eq. (\ref{metric_shear_eq}) (that is, by neglecting the metric shear on the r.h.s.). The vertical black dashed (green dashed) line corresponds to BBN  time  $a_{BBN} \sim 10^{-8}$ (matter-radiation equality). The horizontal  dashed   line corresponds to the constraint set by BBN.   We see that the vector mass should be $m_A \gtrsim 10^{-26} \rm{eV}$ to satisfy the BBN constraints.}
    \label{fig:shear_abundance}
\end{figure}

In Fig. \ref{fig:shear_abundance} we show the numerical evolution of the shear abundance generated by the VFDM. Consistently with the above  analytical approximation, before the field starts oscillating the shear abundance remains constant. The value of the constant can be approximated by the analytical estimate  \cite{Chase:2023puj}:
\begin{equation}
    \Omega_{\sigma} \simeq 4\, \Omega_{DM,0}^2 \Omega_{r,0}^{-3/2}\left(\frac{H_0}{m}\right)\,, \qquad a < a_{osc}\,.
    \label{cond_ini_background_shear}
\end{equation} 
where $\Omega_{DM,0}$ indicates the DM abundance at present time. After the field starts oscillating, the shear has a decaying behaviour which is consistent with  Eq. (\ref{eq_shear_abundance_evolution}).

As shown in \cite{Chase:2023puj}, we can constrain the vector's mass by calculating the shear abundance at Big-Bang Nucleosynthesis (BBN).  
Before the field starts oscillating, although its abundance is well subdominant with respect to radiation, the VFDM affects the background evolution by sourcing the metric shear, leading to   the additional abundance $\Omega_{\sigma}$   (see Eq. \ref{modified_friedmann}). As described in \cite{Barrow:1976rda, Campanelli:2011aa, Akarsu:2019pwn}, BBN leads to  a constraint    on $\Omega_{\sigma}$ at the moment of BBN\footnote{In  \cite{Chase:2023puj} the approximation  $a_{\rm{BBN}} \sim 10^{-8}$ was used to obtain a rough estimate for the bound on $\Omega_{\sigma}$ at BBN, given the bound  $\Omega_{\sigma, 0}\lesssim 10^{-15}$  at $a=1$ taken from \cite{Akarsu:2019pwn}. Actually, our numerical result also gives a rough estimate of the bound. For $m\sim10^{-25}\rm{eV}$ it can be shown that the error in not considering the shear in the vector energy-momentum tensor is about $\sim \%10$. Therefore, a calculation including the shear non-perturbately is needed to obtain a precise bound from BBN. We leave this calculation for future work.}
, $\Omega_{\sigma}\big|_{BBN} \lesssim 10^{-2}$, which we draw in Fig. \ref{fig:shear_abundance} with a black-dashed horizontal line. From the figure we obtain that the vector mass should be $m_A \gtrsim  10^{-26} \rm{eV}$ to satisfy the BBN constraints.

\section{Linear perturbations with VFDM} \label{sec:Vector field perturbations}

In this section we present the implementation of the linear perturbations with VFDM in CLASS. We start by presenting the equations that are implemented in CLASS, and the procedure we use for the implementation.

We work in  synchronous gauge adapted for a Bianchi I background, as defined in Appendix A of \cite{Chase:2023puj}. We work in Fourier space,   
\begin{equation}
    f(\vec{x}) = \int d^3k\,f(\vec{k})\,e^{i\,\vec{k}\cdot\vec{x}}\,,
\end{equation}
where ``$\cdot$'' denotes the product with metric $\gamma_{ij}$ (e.g. $\vec{k}\cdot\vec{x}=k_ix^i$). Here $x^i$ are the comoving coordinates. In Fourier space $k_i$ is constant while $k^i\equiv\gamma^{ij}k_j$ changes with time.

The metric perturbations are given by
\begin{subequations}
\begin{align}
   \delta g_{00} &= \delta g_{0i} = 0\,,\\[7pt]
   \delta g_{ij} &= a^2\left[-2\left(\gamma_{ij}+\frac{\sigma_{ij}}{\mathcal{H}}\right)\, \eta + \hat{k}_i\hat{k}_j (h+6\eta) \right]\,,
\end{align}    
\end{subequations}
where $h$ and $\eta$ are scalar perturbations and $\sigma_{ij}$ is the metric  shear. In this work we neglect vector and tensor perturbations   for simplicity. To describe the perturbations it is convenient to use a mode-dependent base defined as $\{ \hat{e}_1, \hat{e}_2, \hat{e}_3 \}$ where $\hat{e}_3 = \hat{k}$, $\hat{e}_2 = \hat{k} \times \hat{A}$ and $\hat{e}_1 = \hat{k} \times \hat{e}_2$. We can decompose the metric shear in this basis as
\begin{equation}
    \sigma_{ij} = \frac{3}{2} \left( \hat{k}_i \hat{k}_j - \frac{\gamma_{ij}}{3} \right) \sigma_{\parallel} + 2\,\sum_{a=1,2} \sigma_{v_a} \, \hat{k}_{(i}\,\hat{e}_{j)}^a + \sum_{\lambda=+,\times} \sigma_{\lambda} \,\epsilon_{ij}^{\lambda} \,,
\end{equation}
where $\epsilon_{ij}^{+} = \hat{e}_i^1 \,\hat{e}_j^1 - \hat{e}_i^2 \,\hat{e}_j^2$ and $\epsilon_{ij}^{\times} = \hat{e}_i^1\, \hat{e}_j^2 + \hat{e}_i^2 \,\hat{e}_j^1$.

Following the background procedure, we split the vector perturbation dynamics in an equation of motion for the spatial components $\delta \vec{A}$, and a constraint equation for the temporal component $\delta A_0$. For convenience, we   write the spatial components of the vector as a linear combination of the longitudinal mode plus the transverse mode, 
\begin{equation}
    \delta \vec{A} = \delta A_L \,\hat{k} + \delta A_T \,\hat{e}_1.
\end{equation}

We can extract the equations of motion at linear order from Eq. (\ref{general_eq_motion}). We   obtain a constraint equation for the temporal component,
\begin{equation}
    \delta A_0 =  -i \frac{k}{m^2 a^2 + k^2} \left[\delta \dot{A}_L - \frac{1}{2}\dot{A}_L (h + 8 \eta)\right]\,,
\end{equation}
and that the equations of motion for the spatial components can be recasted as   equations for  the longitudinal and the transverse polarizations, respectively, as
\begin{subequations}    
\begin{align}
    &\delta \ddot{A}_L + m^2a^2 \delta A_L - i\,k\, \delta \dot{A}_0 = \frac{1}{2} \dot{A}_L(\dot{h} + 8 \dot{\eta})\,,  \\
    &\delta \ddot{A}_T + \left( m^2a^2 + k^2\right) \delta A_T = -\frac{1}{2} \dot{A}_T(\dot{h} + 4 \dot{\eta}) \,.
\end{align}
\end{subequations}

In the linear regime it is a good approximation to implement the Einstein equations to leading order in $|\sigma_{ij}|\ll \mathcal{H}$ (see  \cite{Chase:2023puj} for details). Following \cite{Chase:2023puj}, the shear tensor has to be included in Einstein $0i$ equation since it is the dominant term at early times far outside the horizon. Then, the Einstein equations to linear order implemented in the code are
\begin{subequations} 
    \begin{align}
        &k^2 \eta - \frac{1}{2} \mathcal{H} \dot{h} = -\frac{a^2}{2 m_P^2} \delta \rho \,,\label{einstein_temporal}\\[7pt] 
        &k^2 \dot{\eta} - \frac{3}{2}k^2 \sigma_{\parallel} \eta = \frac{a^2}{2 m_P^2} (\rho + P) \theta \label{einstein_0i_synch}\,,\\[7pt]
        & \ddot{h} + 2 \mathcal{H} \dot{h} - 2 k^2 \eta  = -\frac{3 a^2}{m_P^2} \delta P\,, \label{einstein_ii_synch} \\[7pt]
        &\ddot{h} + 6 \ddot{\eta} + 2 \mathcal{H} (\dot{h} + 6 \dot{\eta}) - 2 k^2 \eta  = -\frac{3 a^2}{m_P^2} (\rho + P) \delta\Sigma_{\parallel}\,, \label{einstein_ij_synch}
    \end{align}
    \label{einstein_eqs_synch}
\end{subequations}
where $(\rho_A + P_A) \delta\Sigma_{\parallel} = -(\hat{k}_i \hat{k}_j - \frac{\delta_{ij}}{3})(\delta {T^i}_j - \frac{{\delta^i}_j}{3} \delta {T^k}_k)$ and $(\rho_A + P_A) \theta_A = i k^i \delta {T^0}_i$, and where $\delta {T^\mu}_\nu$ denotes the sum of the energy-momentum tensor of all species. The expressions for the fluid variables of the vector field are given in the Appendix of Ref. \ref{appendix_fluid_variables}. The longitudinal projection of the shear tensor $\sigma_{\parallel}$ is written as $\sigma_{\parallel} = 3/2(\cos(\gamma_k)^2 - 1/3) \sigma_A$, where 
\begin{equation}\label{costheta}
    \cos(\gamma_k) = \hat{A}\cdot\hat{k}\,,
\end{equation} 
and is implemented in the background module as the projected Eq. (\ref{metric_shear_eq}) in the direction of the vector field. In the $\sigma_{\parallel}\to 0$ limit the previous set of equations reduce to the usual $\Lambda$CDM Einstein equations, which are already implemented in CLASS. A discussion about the impact of the metric shear in the cosmological observables is given in Appendix  \ref{sec:shear_impact}.

\subsection{Implementation in CLASS} \label{sec:implementation in class}

As for the background, it turns out to be convenient to reduce the equations of motion for the longitudinal and transversal modes to a system of   differential equations of first order. We do this with the following change of variables 
\begin{subequations}
\begin{align}
    \delta A_L =& \sqrt{6} m_P \cos(\gamma_k)\frac{\mathcal{H}}{m} e^{\alpha_L/2} \sqrt{\Omega_A} \sin \left(\frac{\xi_L}{2}\right), \label{change_of_variables_A_xi_alphhaAL}\\[4pt]
    \delta \dot{A}_L =& \sqrt{6} m_P \cos(\gamma_k) a\mathcal{H} e^{\alpha_L/2} \sqrt{\Omega_A} \cos \left(\frac{\xi_L}{2}\right),\label{change_of_variables_A_xi_alphhaBL}\\
    \delta A_T =& \sqrt{6} m_P \sin(\gamma_k)\frac{\mathcal{H}}{m} e^{\alpha_T/2} \sqrt{\Omega_A} \sin \left(\frac{\xi_T}{2}\right), \label{change_of_variables_A_xi_alphhaAT}\\[4pt]
    \delta \dot{A}_T =& \sqrt{6} m_P \sin(\gamma_k) a\mathcal{H} e^{\alpha_T/2} \sqrt{\Omega_A} \cos \left(\frac{\xi_T}{2}\right),
    \label{change_of_variables_A_xi_alphhaBT}
\end{align}\label{change_of_variables_A_xi_alphha}
\end{subequations}
where $\alpha_{L,T}$ and $\xi_{L,T}$ are new variables of linear order in perturbation theory. The new set of equations can be seen in Appendix \ref{appendix:change_of_variables}.

Although the system of equations in Appendix \ref{appendix:change_of_variables} can be solved numerically, it is convenient to take the following change of variables to further simplify the equations:
\begin{subequations}
\begin{align}
    \delta_{L,0} &= 2 \,e^{\alpha_L}\, \sin\left(\frac{\theta - \xi_L}{2}\right) - h - 8 \eta\,, \\
    \delta_{L,1} &= 2 \,e^{\alpha_L}\, \cos\left(\frac{\theta - \xi_L}{2}\right)\,,\\
    \delta_{T,0} &= 2 \,e^{\alpha_T}\, \sin\left(\frac{\theta - \xi_T}{2}\right)\,,\\
    \delta_{T,1} &= 2 \,e^{\alpha_T}\, \cos\left(\frac{\theta - \xi_T}{2}\right)\,.
\end{align}
\label{second_change_of_variables}
\end{subequations}

By deriving the previous expressions and replacing $\alpha^{\prime}_{L,T}$ and $\xi^{\prime}_{L,T}$ with the expression from Appendix \ref{appendix:change_of_variables}, we get a new set of equations of motion:
\begin{align}
    \delta_{L,0}^{\prime} &= -\kappa \left[ \sin\left(\theta\right) + \frac{4(1 + \cos\left(\theta\right))}{4\kappa + y} \right] \delta_{L,0} \label{eq_mov_delta0_L}\\[3pt]
    &\quad+\left[ \sin\left(\theta\right) - \kappa\left(1 + \cos\left(\theta\right) -  \frac{4\sin\left(\theta\right)}{4\kappa + y} \right)\right] \delta_{L,1} \nonumber \\[3pt]
    &\quad- \frac{1}{2}(1-\cos\left(\theta\right)) (h^{\prime} + 8  \eta^{\prime}) \nonumber\,,\\[5pt]
    \delta_{L,1}^{\prime} &= \kappa \left[ 1 - \cos\left(\theta\right) + \frac{4\sin\left(\theta\right)}{4\kappa + y} \right] \delta_{L,0} \label{eq_mov_delta1_L}\\[3pt]
    &\quad+ \left[\cos\left(\theta\right) + \kappa\left(\sin\left(\theta\right) - \frac{4(1-\cos\left(\theta\right))}{4\kappa + y} \right)\right] \delta_{L,1} \nonumber\\[3pt] 
    &\quad- \frac{1}{2} \sin\left(\theta\right) (h^{\prime} + 8  \eta^{\prime}) \nonumber\,,\\[5pt]
    \delta_{T,0}^{\prime} &= -\kappa \sin\left(\theta\right)  \delta_{T,0} + \left[ \sin\left(\theta\right) - \kappa\left(1 + \cos\left(\theta\right)\right) \right]  \delta_{T,1} \nonumber\\[3pt]
    &\quad- \frac{1}{2}(1 + \cos\left(\theta\right)) (h^{\prime} + 4  \eta^{\prime})\,,\label{eq_mov_delta0_T}\\[5pt]
    \delta_{T,1}^{\prime} &= \kappa \left[ 1 - \cos\left(\theta\right)\right] \delta_{T,0} + \left[ \cos\left(\theta\right) + \kappa \sin\left(\theta\right) \right] \delta_{T,1} \nonumber\\[3pt]
    &\quad+ \frac{1}{2} \sin\left(\theta\right) (h^{\prime} + 4  \eta^{\prime})\,,\label{eq_mov_delta1_T}
\end{align}
where $\kappa =  k^2/(2 m a \mathcal{H})$.

We can also express the fluid variables in terms of $\delta_0$ and $\delta_1$. This way, the variables read
\begin{align}
     \delta \rho_A = &\rho_A \frac{\cos(\gamma_k)^2}{4 \kappa + y} \bigg[(2\,\kappa\, (1-\cos\left(\theta\right))  + y) \delta_{L,0} \label{delta_rho}\\[3pt]
     &+ 2 \sin\left(\theta\right) \,\kappa\, \delta_{L,1}\bigg] + \rho_A \sin(\gamma_k)^2 \delta_{T,0}\nonumber\\[3pt]
    &+ \rho_A\,(3 + \cos(2\gamma_k)) \eta\,, \nonumber
\end{align}
\begin{align}
    \delta P_A &= \frac{\rho_A}{3}\frac{\cos(\gamma_k)^2}{4\kappa + y} \bigg[((2\kappa+y)\cos\left(\theta\right))  - 2\kappa)\, \delta_{L,0} \\[3pt]
    &- \sin\left(\theta\right) (2\kappa+y) \delta_{L,1}\bigg] +\frac{\rho_A}{3} \sin(\gamma_k)^2 (\cos(\theta)\delta_{T,0}\nonumber\\[3pt]
    &-\sin(\theta)\delta_{T,1}) + \frac{\rho_A}{3}\,(3 + \cos(2\gamma_k)) \cos\left(\theta\right) \eta\,, \nonumber
\end{align}
\begin{align}
    (\rho_A + &P_A) \theta_{A} =  \rho_A \cos(\gamma_k)^2\,\frac{ a H y \kappa}{4\kappa+y} \bigg[-\sin(\theta) \delta_{L,0} \\[3pt]
    &+ (1-\cos(\theta)) \delta_{L,1}\bigg] + \rho_A \sin(\gamma_k)^2 a H \kappa \nonumber\\[3pt]
    &\times\left[\sin(\theta)  \delta_{T,0} + (1+\cos(\theta))  \,\delta_{T,1}\right]\,,\nonumber
\end{align}
\begin{align}
    (\rho_A &+ P_A) \delta\Sigma_{A,\parallel} = \frac{4}{3}\rho_A \cos(\gamma_k)^2 \frac{1}{4\kappa+y} \times\\[3pt]
    &\times \big[(2 \kappa \left(-1+\cos (\theta)\right)  + \cos (\theta) y) \delta_{L,0} \nonumber\\[3pt]
    &- \sin (\theta) \left(2\kappa+ y\right) \delta_{L,1} \big] + \frac{2}{3}\rho_A \sin(\gamma_k)^2  \nonumber\\[3pt]
    &\times\left[-\cos(\theta)  \delta_{T,0} + \sin(\theta) \delta_{T,1}\right] \nonumber\\[3pt]
    &+\frac{2}{3} \rho_A  (3 + 5 \cos(2\gamma_k))  \cos (\theta)\,\eta\,. \nonumber
\end{align}

The equations of motion and fluid variables of this sections are the ones implemented in the code, plus the correction of Einstein 0i equation (Eq. \ref{einstein_0i_synch}). For every step of the evolution, CLASS numerically solves Eqs. (\ref{eq_mov_delta0_L}), (\ref{eq_mov_delta1_L}), (\ref{eq_mov_delta0_T}) and (\ref{eq_mov_delta1_T}), and then calculates the fluid variables with these solutions. These quantities are then used to source the dark matter sector of the r.h.s of Einstein equations. 

\subsection{Initial conditions}\label{sec:pert_initial_conditions}

In order to calculate the initial conditions for our new variables, we seek for the attractor solutions of the equations of motion far outside the horizon ($k \ll \mathcal{H}$) and deep in the radiation era. In Appendix \cite{Chase:2023puj} some of us showed that the attractor solutions are the adiabatic modes in the sense of Weinberg \cite{Weinberg:2003sw}. In the early universe we can assume a radiation dominated universe where the metric potentials are given by $h = k^2 \frac{\eta_0}{2} (\frac{a}{a_{\rm ini}})^2$ and $\eta = \eta_0(1 - \frac{\alpha_R}{2} k^2 (\frac{a}{a_{\rm ini}})^2)$, with $\alpha_R = \frac{5 + 4 R_{\nu}}{6(15+4R_{\nu})}$ and $\eta_0$ an integration constant (see for example Eq. (96) in \cite{Ma:1995ey}). 

In the radiation era outside the horizon the equations of motion reduces to
\begin{align}
    \delta_{L,0}^{'} &= - \kappa\left[ y + \frac{8}{4\kappa+y}\right]  \delta_{L,0}\, + \\[4pt]
    &\left[y - 2\kappa(1- \frac{2y}{4\kappa+y})\right] \delta_{L,1} - \frac{1}{2} y^2  (1-8\alpha_R)h\,, \nonumber\\[4pt]
    \delta_{L,1}^{'} &= \kappa y\left[\frac{y}{2}+ \frac{4}{4\kappa+y}\right] \delta_{L,0} \\[4pt]
    &+ \left[1 +\kappa y - \frac{2\kappa y^2}{4\kappa+y} \right] \delta_{L,1} - y \left(1-8 \alpha_R \right) h \,, \nonumber\\[4pt]
    \delta_{T,0}^{'} &= - \kappa y\, \delta_{T,0} + (y -2\kappa) \delta_{T,1} - 2 \left(1- 4 \alpha_R \right) h\,, \\[4pt]
    \delta_{T,1}^{'} &= \frac{\kappa y^2}{2} \delta_{T,0} + (1 + \kappa y) \delta_{T,1} + y \left(1-4 \alpha_R \right) h \,,
\end{align}
where we have used that at early times $\theta \sim y \ll 1$ (see Eq. \ref{early_time_solution}), and for the modes outside the horizon we have that $\kappa y \ll 1$.

In the remaining of the paper it is important to differentiate between relativistic and non-relativistic Fourier modes. We refer to a mode as being relativistic or non-relativistic when $k \gg m a$ and $k \ll m a$ respectively. For the lower masses considered in this work, several modes of cosmological interest are still relativistic when CLASS starts evolving them. This will have an impact on the calculation of the initial conditions for the longitudinal variables. Then, we calculate the initial conditions in both the relativistic and non-relativistic regimes.

We can solve the system of equations at early times perturbately. The complete solutions for relativistic and non-relativistic modes to linear order in powers of $\kappa y$ are
\begin{align}
    \delta_{L,0} &= (1 + 4 \frac{\kappa}{y} + 2 \kappa^2 - \frac{3}{2} \kappa y + \frac{5}{2}\kappa^3y) d_L\,,\label{cond_ini_L0}\\[4pt]
    \delta_{L,1} &= -4 \kappa  (1 + \frac{3}{2} \kappa y)d_L\,, \label{cond_ini_L1}\\[4pt]
    \delta_{T,0} &= (1-\frac{1}{2}\kappa y)d_T \,, \label{cond_ini_T0}\\[4pt]
    \delta_{T,1} &= -\frac{1}{10} (5+\kappa y)(1-4\alpha_R) y\,\delta_{\gamma}+ y^2\frac{\kappa}{6} d_T\,, \label{cond_ini_T1}
\end{align}  
where we used the adiabatic initial conditions for the photon energy density $h \sim -\frac{3}{2} \delta_{\gamma}$ (see   \cite{Chase:2023puj} or Eq. (77) in \cite{Ma:1995ey}), and where $d_L$ and $d_T$ are constants of integration. The previous solutions works in both the relativistic and non-relativistic modes.  

To calculate the constants $d_L$ and $d_T$ we impose adiabatic initial conditions for the vector field,
\begin{equation}
    \frac{\delta \rho_A}{\dot{\rho}_A} = \frac{\delta \rho_{\gamma}}{\dot{\rho}_{\gamma}} \quad \longrightarrow \quad    \delta_A = \frac{3}{4} \delta_{\gamma} (1 + w_A).
\end{equation}

Then, we have that at early times $\delta_A = \delta_{\gamma}$. Since $\delta_{\gamma} \sim\mathcal{O}(k^2\tau^2)$, the zero order in the vector overdensity needs to cancel exactly. That is, we need to choose the constants $d_L$ and $d_T$ such that the leading order of $\delta_A$ is $\mathcal{O}(k^2\tau^2)$. By looking at Eq. (\ref{delta_rho}) for early times and on super-horizon scales, we obtain for the constants
\begin{align}
    d_L &= -4 \eta_0\,, \\[4pt]
    d_T &= -2 \eta_0\,.
\end{align}

With this initial conditions we can see that the shear of the vector field $\delta\Sigma_{A,\parallel}$  and the density $\delta_A$ are of $\mathcal{O}(k^2\tau^2)$ at early times; and also that the velocity gradient $\theta_A$ gets exactly cancelled with the term containing the shear tensor on the l.h.s of Eq. (\ref{einstein_0i_synch}). If the shear was not considered in the model, then $\theta_A$ would dominate Eq. (\ref{einstein_0i_synch}) and there would be big infrared contributions from the vector field at early times. In Appendix  \ref{sec:shear_impact} we calculate the error on the matter power spectrum if the shear was not considered in the model.

\section{Evolution of the cosmological perturbations} \label{sec:Cosmological observables}

In this section we focus on the evolution of the cosmological perturbations for fixed directions in space given by $\gamma_k$. We show that for the modes that are relativistic when the field starts oscillating there is an anisotropic imprint in the perturbations and in the cosmological observables. The scale that enters the horizon when the field starts oscillating  corresponds to  the Jeans scale at oscillation, $k_{J,osc} = a_{osc}\sqrt{m H_{osc}}$.  

The calculations are made in synchronous gauge, assuming a negligible but nonzero value of cold dark matter density $(\Omega_{\rm{CDM}} = 10^{-6})$ so that the gauge can be fixed. Since the equations of motion have stiff solutions, we work with time-averaged expressions by replacing the sines and cosines with Eq. (\ref{stiff_trig}).

\subsection{Evolution of VFDM perturbations}\label{sec:evolution_of_VFDM}

In this section we study the VFDM perturbations evolution inside the horizon analytically to understand the observables calculated with CLASS. The perturbations has three characteristic scales ($k$, $m$ and $\mathcal{H}$) which determines different behaviours for the vector field. We separate the analysis in relativistic and non-relativistic modes, and also distinguish between regions where the field is oscillating or not. The modes that are outside the horizon in radiation era were treated in Sec. \ref{sec:pert_initial_conditions}.  

Anisotropies appear in principle for all modes before the field starts oscillating. As mentioned above, after the field  starts oscillating the isotropy theorem does apply, and the equations of motion become isotropic, and therefore independent of $\gamma_k$. Indeed,    the modes that enter the horizon after $a_{osc}$  (which have $k<k_J$ and are already non-relativistics at horizon entry), as shown later, approach to the standard CDM isotropic solution. On the contrary, modes that enter the horizon before $a_{osc}$ are relativistic at horizon entry and behave differently depending on $\gamma_k$. As we show below,  for the latter  it is the homogeneous solution for the density perturbation rather than the particular solution that dominates, leaving to a growth of the  density perturbation and the velocity gradient that depends on $\gamma_k$.  The anisotropic growing behaviour ends at   $a_{osc}$  as expected, but leaves  an anisotropic imprint in the initial conditions for the subsequent isotropic evolution, which shows up for modes whose solution is not dominated by the attractor CDM one,  and is evident for modes with $k>k_J$. 

In the radiation era the metric perturbations $h$ and $\eta$ are determined by the radiation perturbations, and act  as sources of the vector equations of motion. These perturbations   decay   once the mode enter  the horizon, so we will neglect them in the radiation era analysis. In the matter era the perturbation $\eta$ approaches to a constant, so we also neglect it. On the other hand, the metric perturbation $h$ has a growing solution for all the scales in $\Lambda$CDM which follows the CDM density $h^{\prime} \sim 2 \delta_{CDM}^{\prime}$.  We obtain that the vector density turns out to be suppressed for $k > k_J$, and hence in matter domination $h$  is also suppressed. We  then  keep the growing solution of the metric perturbation $h$ in matter era for modes such that $k < k_J$, and neglect  $h$  for  $k > k_J$ and in the radiation. 

In Appendix \ref{sec:jeans_scale} we present an alternative analysis of the Jeans scale in matter era derived using the vector field fluid variables.

\subsubsection{Radiation era for $a<a_{osc}$  inside the horizon}\label{sec:radiation_befor_aosc}

The modes in this regime are relativistic since they are inside the horizon and the field is still not oscillating, so $k \gg \mathcal{H} \gg m a$. In this regime the vector field has an enhancement of the longitudinal modes in comparison with the transverse modes, which amplifies anisotropies leaving an imprint on  observationally relevant quantities such as the matter power spectrum for $k > k_J$. To understand how the anisotropies are amplified we start by considering the equations of motion in the radiation era inside the horizon. We focus on the modes that are relativistic before the field starts oscillating. We thus have that $\kappa y \gg 1$ and $\kappa \gg 1$. To simplify the analysis it is better to consider a new set of variables defined as
\begin{equation}
\begin{cases}
    \tilde{\delta}_{i,0} = - \Upsilon \,\delta_{i,0}\,, \\[4pt]
    \tilde{\delta}_{i,1} = \kappa\, \delta_{i,1} \,.
\end{cases}
\end{equation}
where $\Upsilon = \kappa y/2$ and where $i = L, T$. The equations of motion in the $\Upsilon \gg 1$ regime can be written as
\begin{align}
    \frac{{d \tilde\delta}_{i,0}}
{d\Upsilon} &\simeq - \tilde{\delta}_{i,0} + \tilde{\delta}_{i,1}, \\[4pt]
    \frac{d\tilde{\delta}_{i,1}} {d\Upsilon}&= - \tilde{\delta}_{i,0} + \tilde{\delta}_{i,1},
\end{align}
where we have neglected the metric potentials since they have decaying modes in the radiation era inside the horizon.

We can solve the previous system and fix the constants of integration by demanding continuity when the modes enter the horizon. That is, by matching the solutions with the initial conditions (Eqs. \ref{cond_ini_L0}, \ref{cond_ini_L1}, \ref{cond_ini_T0} and \ref{cond_ini_T1}). We then obtain
\begin{align}
    \tilde{\delta}_{L,0} &= -8 \kappa^2(1 + \Upsilon) \eta_0\,, \\[4pt]
    \tilde{\delta}_{L,1} &= 8 \kappa^2 (2-\Upsilon)\, \eta_0\,,\\[4pt]
    \tilde{\delta}_{T,0} &= 2\eta_0 \Upsilon\,, \\[4pt]
    \tilde{\delta}_{T,1} &= 2 (1 + \Upsilon)\, \eta_0\,.
\end{align}

By inserting the previous solutions into the fluid variables we see that the evolution is anisotropic at leading order,
\begin{align}
    \delta_A &= \left[28 \Upsilon\,\cos(\gamma_k)^2 - 2 \sin(\gamma_k)^2 \right]\eta_0\,, \\[4pt]
    \theta_A &= - \mathcal{H} \left[6 \Upsilon^2 \cos(\gamma_k)^2 + 4 \sin(\gamma_k)^2 \right]\eta_0\,.
\end{align}

In particular, we observe that the longitudinal modes ($\gamma_k = 0$) grow  much faster than the transversal modes ($\gamma_k = \frac{\pi}{2}$) in this regime since $\Upsilon\gg1$, both for $\delta_A$ and $\theta_A$. This behavior can be seen in the numerical solutions of the bottom panels of Figs. \ref{fig:delta_angles} and \ref{fig:theta_angles}, when $a_{\ast} < a < a_{eq}$. 

\subsubsection{Radiation era for $a>a_{osc}$  inside the horizon} \label{sec:radiation_after_aosc}
After the field starts oscillating and before the matter era, it is  convenient  to analyze separately relativistic and non-relativistic modes. Since the field is oscillating, the total equation of state is $w_T \sim w_A \sim 0$ and the shear tensor averages to zero, so we neglect $\sigma_{ij}$ in the equations.
 As we show next, the non-relativistic modes approach to an isotropic solution that correspond to an attractor solution driven by the metric perturbations. However, the relativistic modes still present anisotropies, due to both  the initial conditions and because the transverse and longitudinal modes behave differently also   in this regime. The transverse modes oscillate  with a constant amplitude, while the longitudinal modes oscillate  with a decaying amplitude. This can be seen from the equations of motion in the $\kappa\gg y$ regime. Also, since most of the modes of cosmological interest are inside the horizon in this regime, we have that $\kappa y \gg 1$. Then, the equations of motion reduce  to
\begin{align}
    \delta_{L,0}^{\prime} &\simeq - \delta_{L,0} - \kappa\, \delta_{L,1} - \frac{1}{2} (h^{\prime} + 8 \eta^{\prime})\,, \label{delta_sys_rad_inside_horizon_L0}\\
    \delta_{L,1}^{\prime} &\simeq \kappa \,\delta_{L,0} - \delta_{L,1}\,,\label{delta_sys_rad_inside_horizon_L1}\\[7pt]
    \delta_{T,0}^{\prime} &\simeq - \kappa \,\delta_{T,1} - \frac{1}{2} (h^{\prime} + 4 \eta^{\prime})\,, \label{delta_sys_rad_inside_horizon_T0}\\
    \delta_{T,1}^{\prime} &\simeq \kappa \,\delta_{T,0} \,.\label{delta_sys_rad_inside_horizon_T1}
\end{align}

In the radiation era the metric perturbations $h$ and $\eta$ have decaying solutions once the mode enters the horizon.  In order to solve the homogeneous part  of Eqs. (\ref{delta_sys_rad_inside_horizon_L0}-\ref{delta_sys_rad_inside_horizon_T1})  we combine the equations into two second order equations,  
\begin{align}
    \delta^{\prime\prime}_{L, i} + 2 \delta^{\prime}_{L, i} + \kappa^2 \delta_{L, i} &\simeq 0\,,\\[4pt]
    \delta^{\prime\prime}_{T, i} + \kappa^2 \delta_{T, i} &\simeq 0\,,
\end{align}

\noindent where the subscript i indicates either $1$ or $2$. These   equations can be solved by
\begin{align}
    \delta_{L, i} &\sim e^{-N} \cos(\kappa N)\,, \\[4pt]
    \delta_{T, i} &\sim \cos(\kappa N)\,,
\end{align}

\noindent where $N = \log(a)$. With the field solutions we can now write the vector density and velocity gradient as
\begin{align}
    \delta_A &\sim \cos(\kappa N) \left[\frac{1}{2}e^{-N} \cos^2(\gamma_k) + \sin^2(\gamma_k) \right]\,,\\
    \theta_A &\sim \mathcal{H} \cos(\kappa N) \left[\frac{y}{2}e^{-N} \cos^2(\gamma_k) + \kappa \sin^2(\gamma_k) \right]\,.
\end{align}

\begin{figure}[th]
    \includegraphics[width=0.48\textwidth]{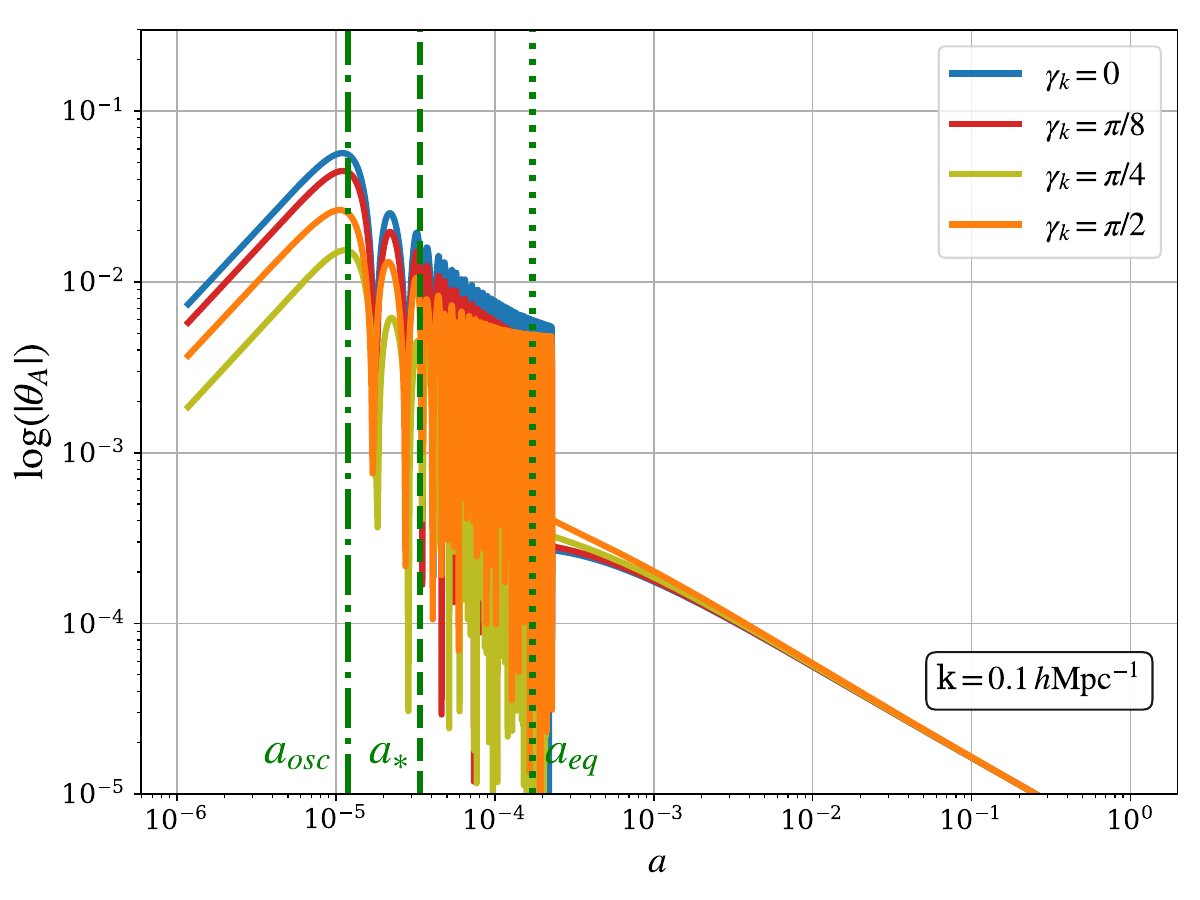}
    \includegraphics[width=0.48\textwidth]{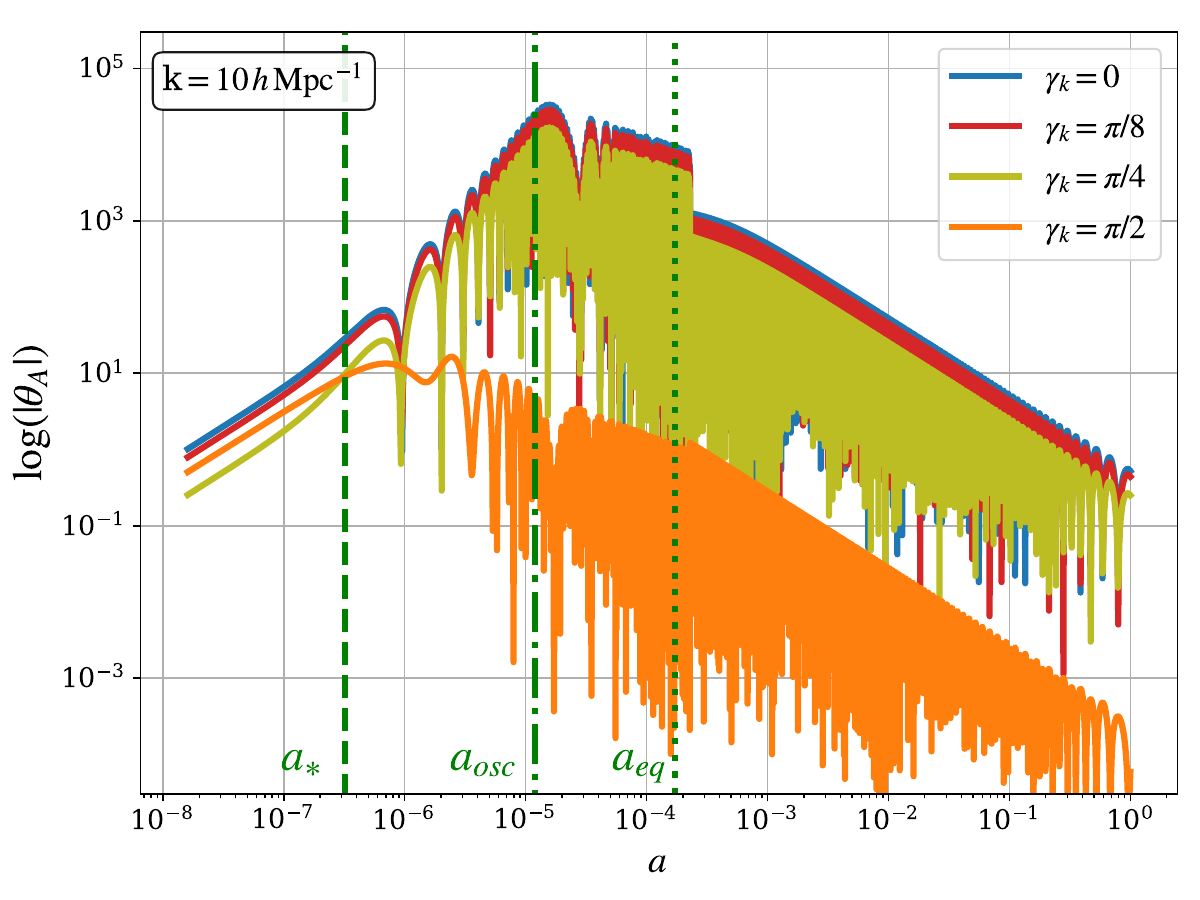}
\caption{Numerical solutions for the VFDM velocity gradient $\theta_A$, for $m = 10^{-25} \rm{eV}$ and different angles $\gamma_k$. The vertical dashed lines indicates horizon entry of the mode defined as $k=\mathcal{H}(a_{\ast})$, and vertical dashed-dot lines indicates the time of oscillation defined as $m a_{osc}=\mathcal{H}(a_{osc})$. We can see that the modes that enter the horizon after the time of oscillation has no anisotropies in the subsequent evolution. On the other hand, the modes that enter the horizon before the time of oscillation has an angle dependence that leaves an anisotropic impint in the later evolution. For the mass considered here the mode that enters the horizon when the field starts oscillating is $k_J \sim 0.27 \,h \, \rm{Mpc}^{-1}$.}
\label{fig:theta_angles}
\end{figure}

The behaviour in this regime  can be seen in  Figs. \ref{fig:delta_angles} and \ref{fig:theta_angles} for $a_{osc} < a < a_{eq}$. The importance of the source terms in  Eqs. (\ref{delta_sys_rad_inside_horizon_L0}-\ref{delta_sys_rad_inside_horizon_T1}) depends on the initial relative amplitude.       
For the modes with $k < k_J$,  it can be seen in the top panels of Figs. \ref{fig:delta_angles} that the vector follow an attractor solution.  This regime corresponds to the modes that are non-relativistic once the field starts oscillating, so we can approximate $y \gg \kappa$. 

Then, by replacing the above equations of motion in the time derivative of the vector's density we obtain the usual continuity equation for CDM, namely $\delta^{\prime} = - h^{\prime}/2 + \mathcal{O}(\kappa)$, where we neglected the decaying and  oscillatory terms with respect to the source. We see that for the modes $k < k_J$ when the field starts oscillating, the density perturbation has an attractor solution which is isotropic.   
The same behaviour continues in the matter domination era, as we show in the following section. This can be also seen in the top panels of Figs. \ref{fig:delta_angles}, for $a_{eq} < a$.

\subsubsection{Matter era inside the horizon}
Now we focus on the perturbations in the matter era inside the horizon. This regime is important to   understand the shape of the matter power spectrums with VFDM.  As we show, VFDM behaves as CDM at large scales, while $\delta_A$ oscillates on small scales, leading to the characteristic suppression in the power spectrums.

\begin{figure}[th]
    \centering 
    \includegraphics[width=0.48\textwidth]{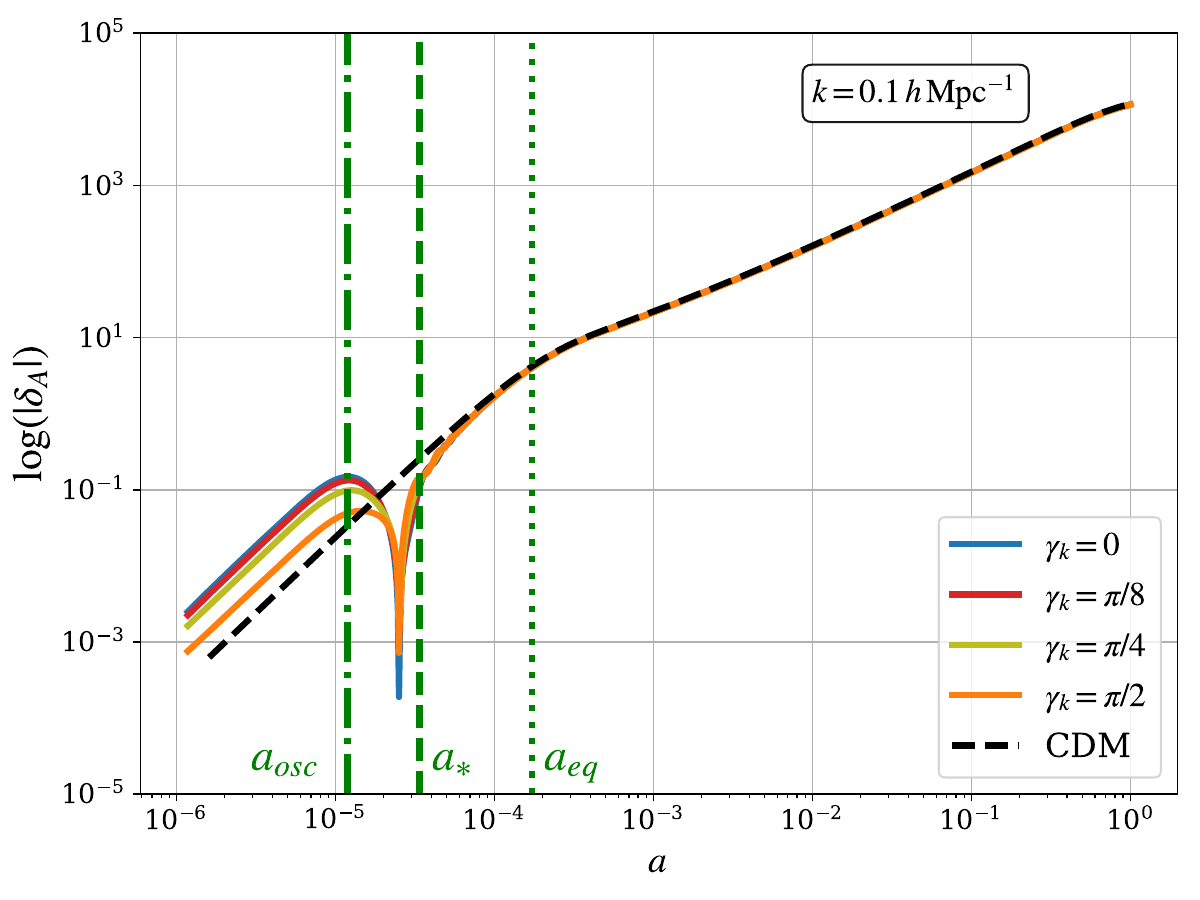}
    \includegraphics[width=0.48\textwidth]{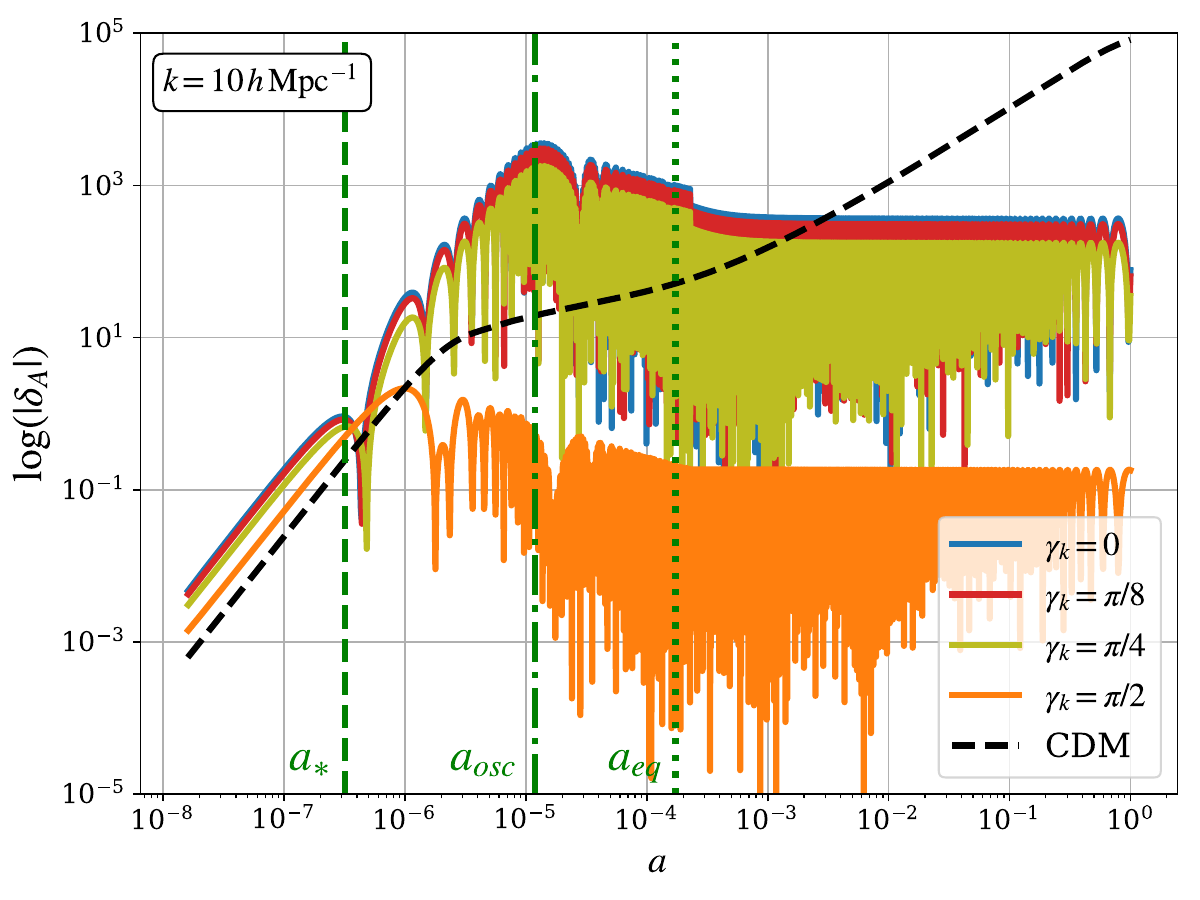}
\caption{Numerical solutions for the VFDM overdensity $\delta_A$, for $m = 10^{-25} \rm{eV}$ and different angles $\gamma_k$. Solid lines indicate the vector's overdensity, while dashed black lines indicate CDM overdensities. The vertical dashed (dashed-dotted) lines indicate horizon entry of the modes (the time of oscillation)  defined as $k=\mathcal{H}(a_{\ast})$, ($m a_{osc}=\mathcal{H}(a_{osc})$), and the vertical dotted lines correspond to the equality between matter and radiation abundances. We can see that the modes that enter the horizon after the time of oscillation evolve towards  isotropic attractor solutions that resemble their CDM counterpart. On the other hand, the modes that enter the horizon before the time of oscillation have an angle dependence.   Anisotropies are amplified between $a_{\ast}$ and  $a_{osc}$ and are imprinted in the subsequent  evolution. For the mass considered here the modes that enter the horizon when the field starts oscillating correspond to $k=k_J \sim 0.27 \,h\,\rm{Mpc}^{-1}$.}
\label{fig:delta_angles}
\end{figure}

To solve for the vector variables in the regime previously described, we need to consider the equations of motion along Einstein equations since the vector field is the main source on the r.h.s in this regime. To solve for the vector overdensity we need to calculate the metric perturbations from Einstein equations as functions of the VFDM fluid variables. We can do this with Eqs. (\ref{einstein_temporal}), (\ref{einstein_0i_synch}) and (\ref{einstein_ii_synch}) by setting $\sigma_{\parallel} = 0$ and by assuming that the r.h.s of the equations are only given by the vector fluid variables.

We can write $\eta^{\prime}$ in terms of the vector field through Eq. (\ref{einstein_0i_synch}) by neglecting the contribution from the other species. However, metric perturbation $h^{\prime}$ cannot be decoupled from $\eta$ in Einstein's equations\footnote{Using the continuity equation  for a general fluid (Eq. (29) in \cite{Ma:1995ey}), solving for $\eta$ gives a trivial equation.}, so we will assume a functional form for the different regimes. 

For large scales, the VFDM mass power spectrum follows the $\Lambda$CDM spectrum, so in this regime we can assume a CDM behaviour for the vector field. Then, we can use the CDM growing mode as a source of Einstein's equations  $\delta_{cdm} \propto a$ and $\theta_{cdm} = 0$. Inserting the CDM variables in Eqs. (\ref{einstein_temporal}) and (\ref{einstein_0i_synch}) we obtain for the metric perturbations $\eta = \eta_{eq} = const$ and $h = h_{eq} e^{N}$, where $h_0 = h_0(k)$. Then, we can turn the equations of motion in a single equation for $\delta_0$, the relevant variable for $\delta_A$ when the field is oscillating, given by
\begin{equation}
    \delta_{i,0}^{\prime \prime} + \frac{1}{2} \delta_{i,0}^{\prime} + \kappa^2 \delta_{i,0} = -h\,,
\end{equation}
where i runs for L and T.

Solving for $\delta_0$ we obtain for $\Lambda$CDM modes
\begin{align}
    &\delta_{i,0} = C_1 \cos(\kappa) - C_2 \sin(\kappa)- \frac{1}{2} h \\[4pt]
    &\times [1 + \kappa^2  \cos(\kappa)  \text{Ci}(\kappa) + \kappa^2 \sin(\kappa) \text{Si}(\kappa)]\,,\nonumber
\end{align}
where $C_1$ and $C_2$ are constants of integration, and Ci and Si are the cosine and sine integrals respectively. Since $\kappa = \kappa_0\, e^{-N/2}$ in matter epoch, the CDM $-\tfrac{1}{2} h$ term dominates at late times. 
We see that for modes where $k \ll k_J$ the growing solution of $h$ dominates. For modes where $k \gg k_J$ when entering matter epoch, we expect oscillations till $k \lesssim k_J$.

In this regime we can write the energy density as $\delta_A \sim \cos(\gamma_k)^2 \delta_{L,0} + \sin(\gamma_k)^2 \delta_{T,0}$. Then, by replacing the field solutions for modes such that $k < k_J$, we obtain an isotropic energy density. Also, we recover the usual continuity equation for cold dark matter,  $\delta^{\prime}_A = -h^{\prime}/2$, and obtain an attractor solution independent of the angle $\gamma_k$. This behaviour can be seen in the top panel of Fig. \ref{fig:delta_angles}. 

For suppressed modes on matter epoch we can solve the potentials using Einstein's equations without sources, since we expect that no VFDM overdensity is generated in these regime. It can be shown that the only consistent solution of Eqs. (\ref{einstein_temporal}), (\ref{einstein_0i_synch}) and (\ref{einstein_ii_synch}) in vacuum are $h = const$ and $\eta=0$. However, since matter perturbation are suppressed but non-zero in this limit we will assume an exponential decay for the potentials, so we can parameterize metric perturbations as $h^{\prime} = h_0 e^{-\beta N}$, where $h_0 = h_0(k)$ and $\beta > 0$. The homogeneous solution for the transverse mode are the same as in the previous regime, so we focus on the longitudinal variables. Then, we can write an equation of motion for $\delta_{i,L}$ as
\begin{equation}
    \delta_{L, i}^{\prime \prime} + \frac{5}{2} \delta_{L, i}^{\prime} + \frac{\kappa^2}{4} \delta_{L, i} = -\frac{5}{4}h\,.
\end{equation}

For simplicity we will consider the case where $\beta = 1$, so we can write the solutions as
\begin{equation}
    \delta_{L, i} =  C_1 \cos \left(\kappa \right) - C_2 \sin \left(\kappa\right) + const.
\end{equation}

Since $\kappa \gg 1$ for the suppressed modes we have high oscillations in $\delta_A$ and no growing mode, as can be seen in the lower panel of Fig. \ref{fig:delta_angles}. This oscillations continue until $k \sim k_J$, and are reflected in the mass power spectrum as a suppression for modes with $k \gg k_J$.

\subsection{Power spectrums}
In VFDM models we can define a direction dependent matter power spectrum as 
\begin{equation}
    \langle\delta(\tau,\vec{k})\delta^*(\tau,\vec{k'})\rangle=(2\pi)^3 \delta^3(\vec{k}-\vec{k'})P({k},{{\gamma_k}}).
\end{equation}

In this section we present the results for the   matter power spectrums obtained by evaluating $P({k},{{\gamma_k}})$ at  specific values of ${\gamma_k}$, calculated with the modified version of CLASS. All plots assume Planck best fit parameters \cite{Planck:2018vyg}.
As we showed in Sec. \ref{sec:evolution_of_VFDM}, the VFDM model presents anisotropies at scales smaller than the Jeans scale ($k > k_J$) in the radiation era. In this section we show that this anisotropies leave an imprint on the power spectrums at such scales.
We compare the power spectrums in VFDM models with the ones for $\Lambda$CDM  and   SFDM models, calculated respectively with CLASS and a CLASS V3.2 code modification based on \cite{Urena-Lopez:2015gur} named \textit{class.SFDM}\footnote{\url{https://github.com/classULDM/class.SFDM}}.

\begin{figure}[th]
    \includegraphics[width=0.48\textwidth]{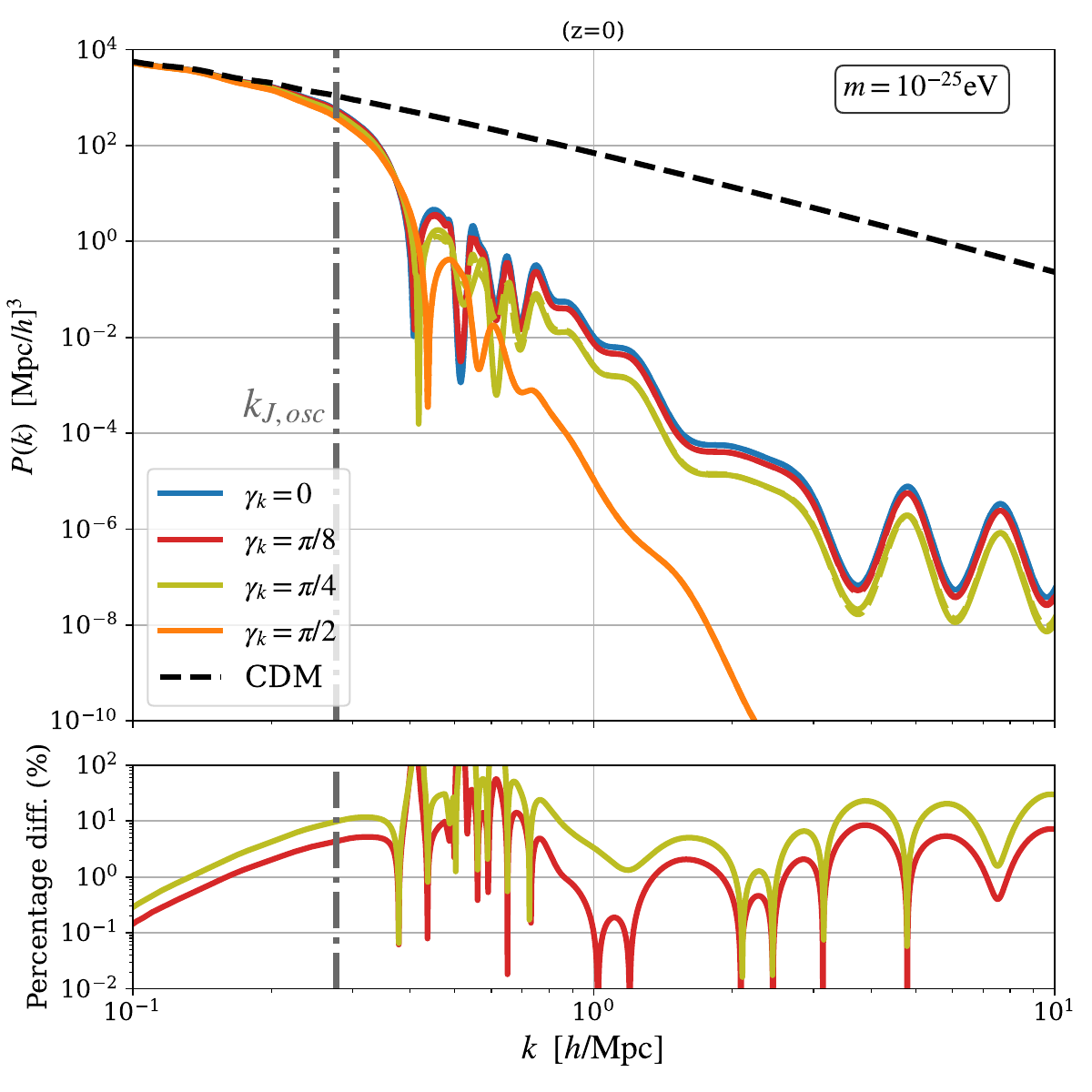}
    \caption{In solid lines we present the matter power spectrums of the VFDM model with $m = 10^{-25}\rm{eV}$ at $z=0$, for different orientations of the Fourier modes with respect to the  background field, parameterized with the angle $\gamma_k$ such that $\cos(\gamma_k) = \hat{k}\cdot\hat{A}$. The vertical dash-dot line corresponds to the Jeans scale $k_J$ at radiation era.  We can see that for  $k\gtrsim k_J$  the power spectrum  is anisotropic. In colored dashed lines we show the parametrization of the power spectrum as a function of $\gamma_k$ given in Eq. (\ref{fit_Pk}). In the lower panel we show the percentage difference of the different orientations with respect to their respective parametrization in Eq. (\ref{fit_Pk}).}
    \label{fig:Pk_m_-25}
\end{figure}

As an example, in Fig. \ref{fig:Pk_m_-25} we present the results for the VFDM matter power spectrum  for different values of $\gamma_k$ and a particular value of the mass $m=10^{-25}{\rm eV}$ at redshift zero ($z=0$). In dashed black line we present the one for the standard  CDM  model. For modes greater than the Jeans scale ($k < k_J$) the VFDM matter power spectrum is isotropic up to a $\leq 10\%$. However, for   the vector model, scales smaller than the Jeans scale ($k > k_J$)   have a suppression in the matter power spectrum, which is anisotropic. In particular, we find that the power spectrum can be parameterized as (see the colored dashed curves in Fig. \ref{fig:Pk_m_-25})
\begin{equation}
    P(k, \gamma_k) = P(k, \frac{\pi}{2}) \left(1 - g(k)\cos(\gamma_k)^4\right)\,,
    \label{fit_Pk}
\end{equation}
where $g(k) = 1 - P(k, 0)/P(k, \frac{\pi}{2})$.

The matter power spectrum in the standard  CDM  case is  isotropic  and is consistent with a great amount of observational data \cite{Chabanier:2019eai}. Therefore, for the VFDM model to be consistent with   observations, we expect that the power spectrums do not change significantly with the angle $\gamma_k$, at least on the largest relevant scales for such  observational data.  In order to quantify the comparison between the different models and the direction dependence of the VFDM power spectrums we define  the percentage differences  among two spectrums as $\% {\rm Diff} = 100\times(P_A-P_B)/P_B$, where the spectrum  $P_B$ corresponds to  the VFDM model with $\gamma_k = 0$ (unless said otherwise).

\begin{figure*}[th]
\begin{minipage}{0.49\linewidth}
    \includegraphics[width=\linewidth]{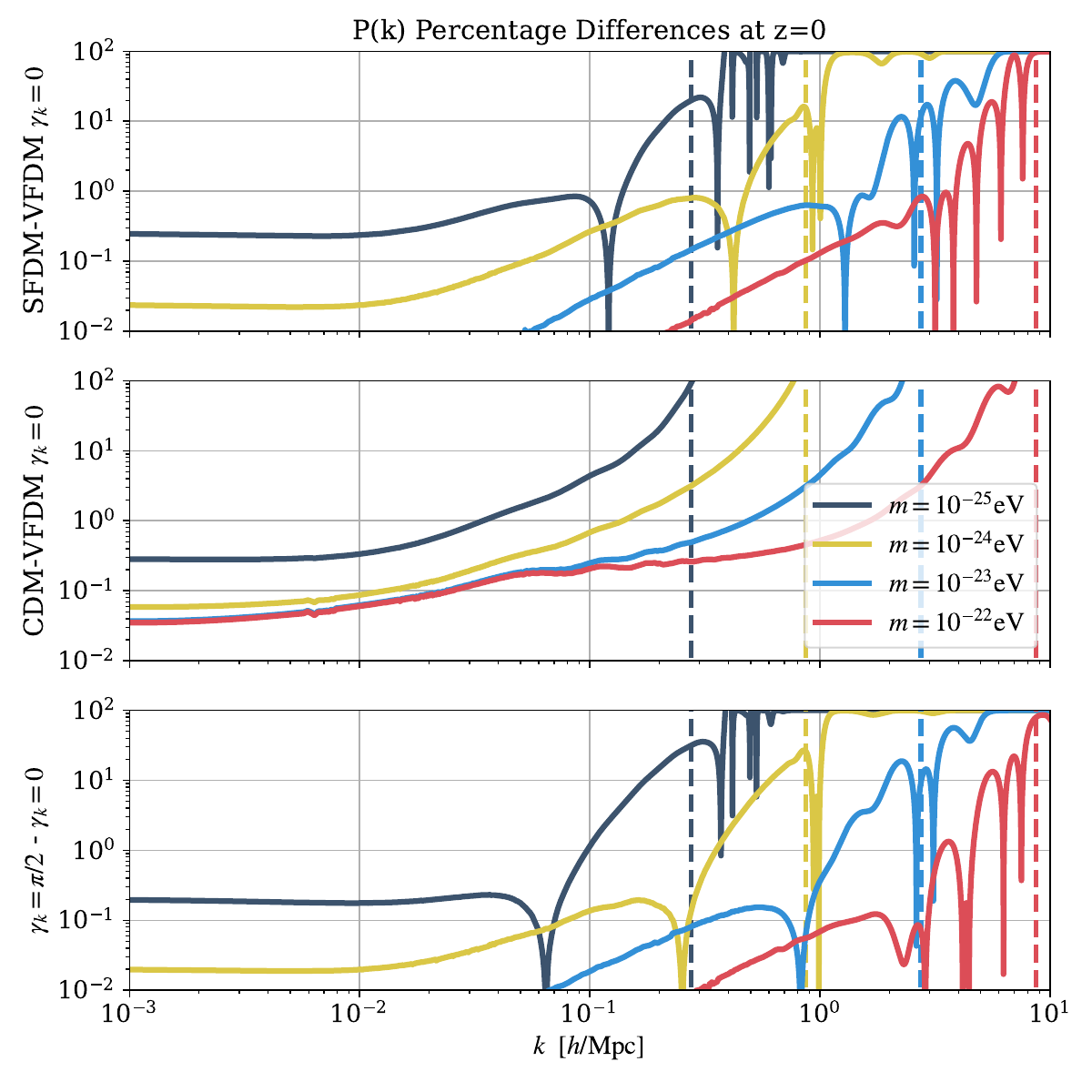}
\end{minipage}
    \hfill
\begin{minipage}{0.49\linewidth}
    \includegraphics[width=\linewidth]{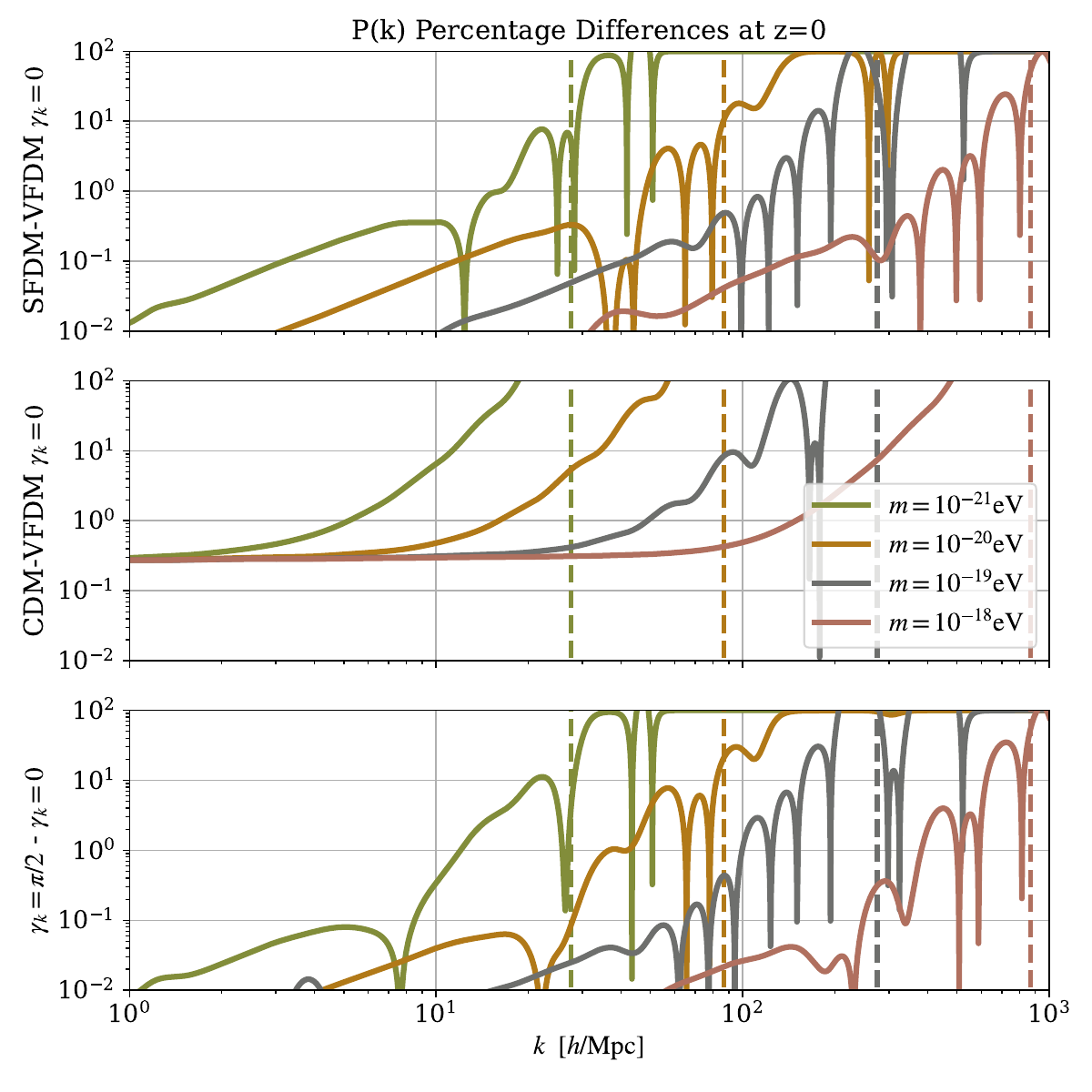}
\end{minipage}
    \caption{Percentage  differences between the power spectrums for  SFDM model (top panels), $\rm{CDM}$ (mid panels) and  VFDM with $\gamma_k = \pi/2$ (bottom panels), with respect to  that for VFDM with $\gamma_k = 0$. The vertical dashed lines indicate the Jeans scale at radiation for each mass. We can see that the percentage difference between the models is $< 1\%$ for $k \ll k_J$. However, for $k \sim k_J$ we obtain differences of around $10\%$ or larger depending on the plot and value of the mass parameter.}
    \label{fig:Pk_Porcentual_errors6}
\end{figure*}

In Fig. \ref{fig:Pk_Porcentual_errors6} we can see the percentage differences in the matter power spectrum between the VFDM model with $\gamma_k = 0$ and $\rm{\Lambda CDM}$, the SFDM model and the VFDM with $\gamma_k = \pi/2$. For modes such that $k \ll k_J$ we have that the differences between the models is $< 1\%$ for $m \gtrsim 10^{-25}\rm{eV}$. However, for modes  with $k> k_J$  we obtain differences of about tens of percent or larger depending on the quantity and the value of the mass of the field. In particular, differences between the power spectrums for longitudinal and transversal modes are  $> 10\%$ for $k\sim k_J$ generically for all the masses plotted. 

As mentioned in the introduction the predictions  for the  large-scale cosmological perturbations in SFDM models have been broadly studied and cosmological data have been used to test different mass ranges of the field \cite{Hlozek:2014lca, Urena-Lopez:2015gur, Hlozek:2017zzf, Lague:2021frh,Rogers:2020ltq}. These observables are sensitive to the typical short scale suppression at $k\sim k_J$ of the matter power spectrum in SFDM models. So far current studies indicate such suppression is absent. To date, the strongest bound is given by Lyman-$\alpha$ forest \cite{Rogers:2020ltq}, which gives $m> 2\times 10^{-20} {\rm{eV}}$.

In this paper we show that a suppression of the matter power spectrum  similar to the one produced in SFDM models  also characterise the shape of the matter power spectrum in VFDM models, for the same Jeans scale. Therefore, we expect that the absence of a detection of such characteristic suppression can also be used  to set a bound on the mass of the VFDM candidates of about the same order, by using the same current data. However,   due to the  fact that the suppression is direction-dependent for VFDM at $k\gtrsim k_J$,  the
comparison between the model predictions and data becomes more involved  and more work is needed to obtain precise  bounds on the mass for VFDM. If such suppression is inferred in the future it could be used as a benchmark to assess whether a model with non-vanishing spin is preferred by the data or not.

Looking into the future, the next frontier are the observations of the redshifted 21 $\rm{cm}$ line of neutral hydrogen. Forecasts in Ref \cite{Munoz:2019hjh} analyze the precision with which 21-cm experiments could probe the matter power spectrum during cosmic down. For example  a global-signal experiment could measure the amplitude of the power spectrum integrated over $k=(40-80)\rm{Mpc}^{-1}$ with a precision of tens of percent while a fluctuation experiment could constrain the power spectrum  to a similar accuracy in bins covering $k=(40-60)\rm{Mpc}^{-1}$  and $k=(60-80)\rm{Mpc}^{-1}$  even without astrophysical priors. Therefore,  forecasts indicates that the  21-cm data could be a powerful probe to distinguish  different  DM candidates such as SFDM and VDFM, among others. In the case of the VDFM, it would be crucial to see a direction dependence of the observable. 

\begin{figure}[th]
    \includegraphics[width=\linewidth]{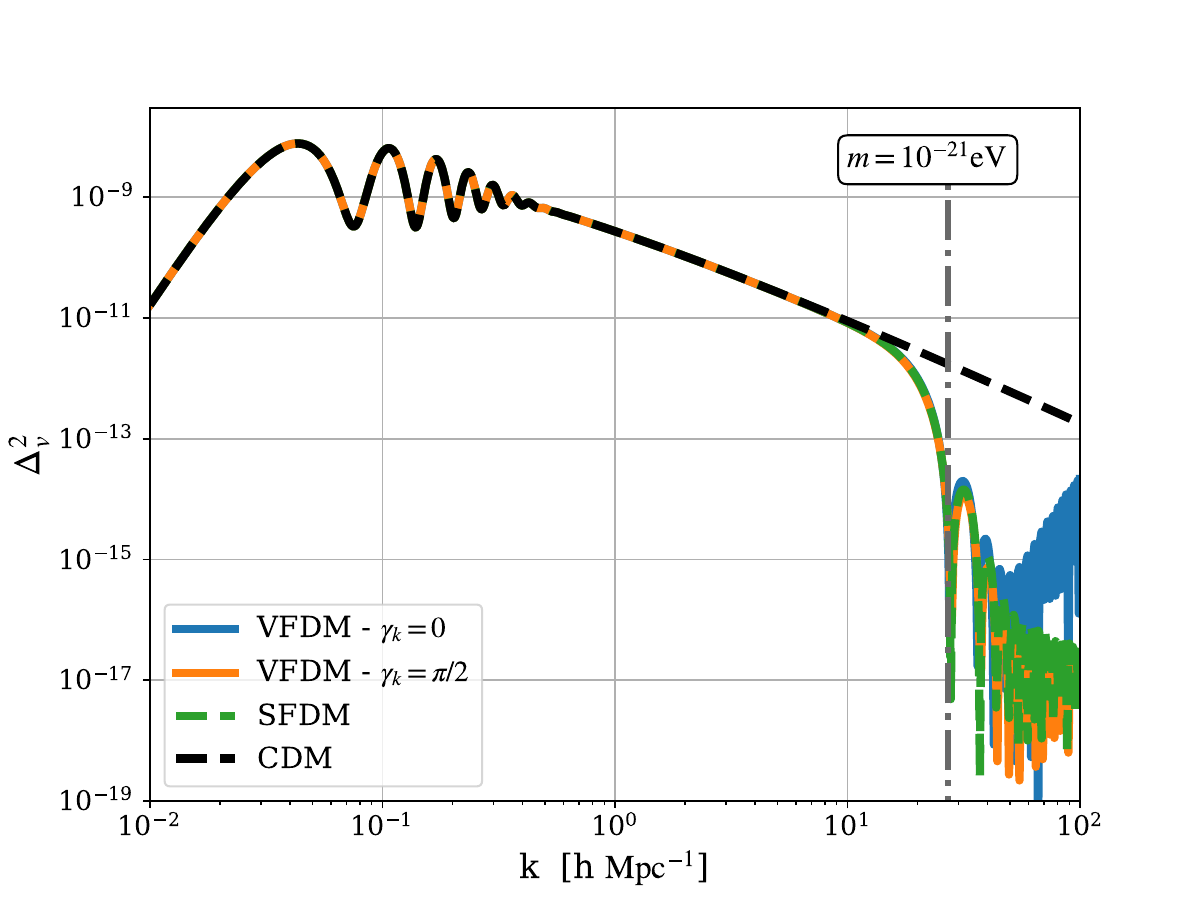}
    \caption{Velocity gradient power spectrums   defined in (\ref{eq:velocity_power_spectrum}) at $z=1020$ for $m = 10^{-21}\rm{eV}$. The green dashed lines indicate the velocity  power spectrum   calculated for the SFDM case, for the same mass. For $k< k_J$   we see agreement between both models regardless the orientation  of the Fourier modes. However, for $k\gtrsim k_J$  there are  anisotropies that can be  understood from the  dependence on $\gamma_k$  of the  evolution of $\theta_A$ given in Fig. \ref{fig:theta_angles}.}
    \label{fig:invariant}
\end{figure}

\begin{figure}[th]
    \includegraphics[width=\linewidth]{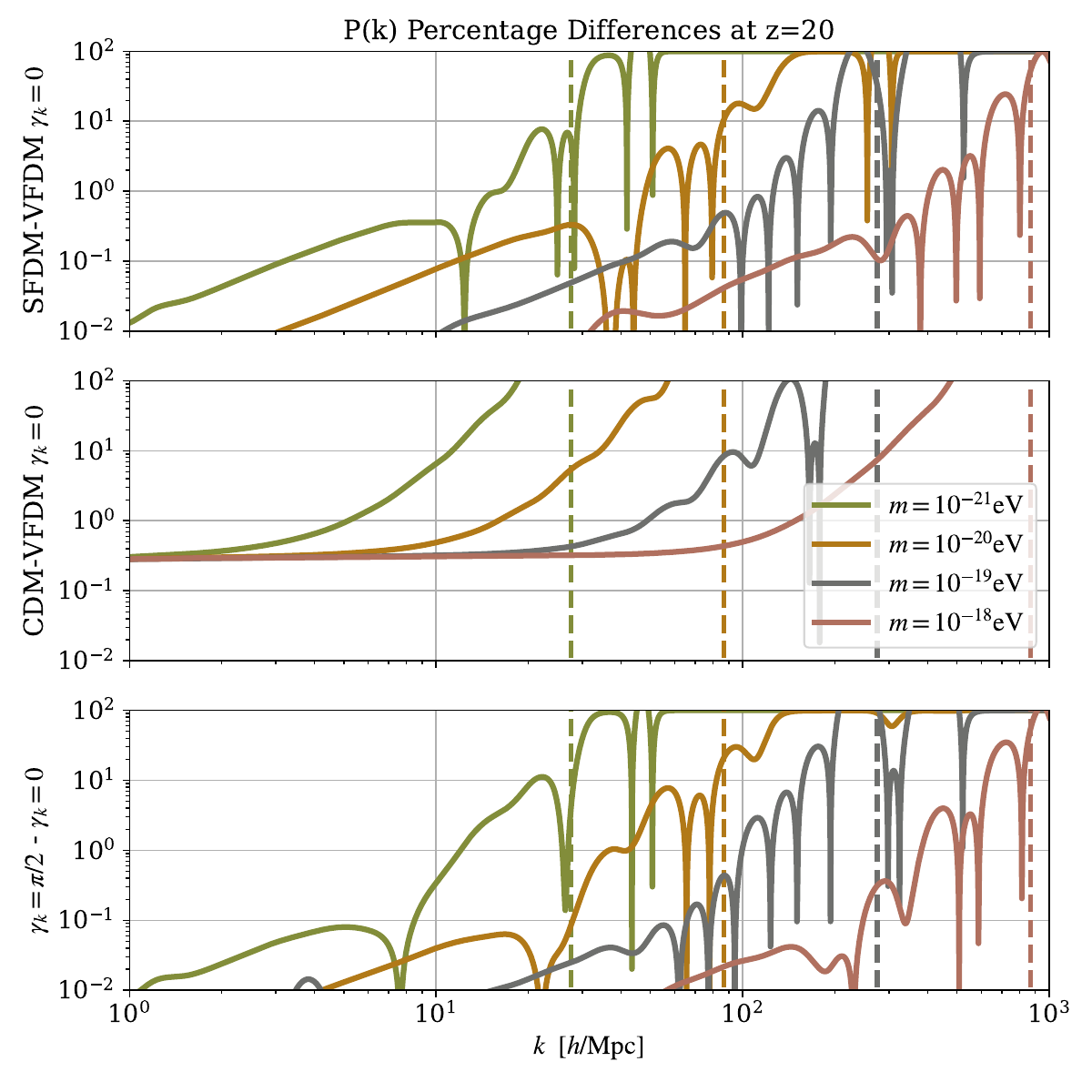}
    \caption{Percentage  differences between the power spectrums for  SFDM model (top panel), $\rm{CDM}$ (mid panel) and  VFDM with $\gamma_k = \pi/2$ (bottom panel), with respect to that for VFDM with $\gamma_k = 0$ at $z=20$. The vertical dashed lines indicate the Jeans scale at radiation for each mass. We can see that the percentage difference between the models is $< 1\%$ for $k \ll k_J$. However, for $k \sim k_J$ we obtain differences  $> 10\%$ for the upper and bottom plots and  $> 100\%$ for the one in the middle.}
    \label{fig:Pk_Porcentual_errors}
\end{figure}

Forecasts such as those in \cite{Munoz:2019hjh} generally assume that the  relative velocity between CDM and baryons power spectrum around recombination is the same for any DM candidate.  
On the other hand, as emphasised  in \cite{Marsh:2015daa} for SFDM, the relative velocity between SFDM and baryons power spectrum  at recombination (around $z=1020$) is a relevant output of the Boltzmann codes to study the non-linear physics involved in the prediction of the 21-cm power spectrum. The  reason is a well-known  suppression  of star formation in the first structures at high redshift produced by  such relative velocity  (see for instance \cite{Fialkov:2014rba} and references therein). This affects the 21 cm signal during Cosmic Dawn. 
 The variance of the relative velocity between baryons and DM, $\vec{v}_{b,dm} = \vec{v}_{b} - \vec{v}_{dm}$, is defined as
\begin{equation}
    \langle \vec{v}_{b,dm}^{\,2} \rangle = \int \frac{dk}{k} \Delta^{2}_{v}(k) \, .
    \label{eq:velocity_power_spectrum_integral}
\end{equation}
Here $\Delta_{v}(k)$ is the relative velocity power spectrum which is defined by
\begin{equation}
    \Delta_{v}^2(k) = \Delta_{\zeta}^2(k) \left( \frac{\theta_b - \theta_{dm}}{k} \right)^2 \, ,
    \label{eq:velocity_power_spectrum}
\end{equation}
where $\Delta_{\zeta}^2(k)$ is the primordial scalar power spectrum, $\theta_{dm}$ is the velocity divergence of DM and $\theta_b$ is the baryon velocity divergence. 
In Fig. \ref{fig:invariant} we show the relative velocity power spectrum at around recombination  ($z=1020$) for a mass of the vector field $m =  10^{-21}\rm{eV}$, for $\gamma_k=0$ (blue) and $\gamma_k=\pi/2$ (orange). Also, in green dashed  line and black dashed we show the SFDM and CDM relative velocity power spectrums.  As for the matter power spectrum, the relative VFDM-baryon power spectrum is indistinguishable from that of  CDM or SFDM for $k\ll k_J$, with differences appearing only for modes with $k\gtrsim k_J$. As shown in Fig. \ref{fig:Pk_Porcentual_errors} we see that the power spectrum is anisotropic for  modes with  $k\gtrsim k_J$. These anisotropies are produced by the enhancement in the fluid variables in the longitudinal modes with respect to the transverse modes (see Sec. \ref{sec:evolution_of_VFDM} for a discussion of this enhancement).

In Ref. \cite{Hotinli:2021vxg}, deepening in the study of the relative velocity effect in SFDM initiated in \cite{Marsh:2015daa},  the authors studied the velocity acoustic oscillations (VAOs) \cite{Munoz:2019fkt} in the large-scale 21-cm power spectrum.  The VAO  features are interpreted as the result of the modulation of short scales  ($k\sim 10-10^3 \,\rm{Mpc}^{-1}$)  by DM–baryon relative velocities of the long-wavelengths ($k\sim 0.1 \, \rm{Mpc}^{-1}$).
For SFDM, the effect on the 21 cm power spectrum was studied in \cite{Marsh:2015daa, Hotinli:2021vxg},  showing that the VAO features  are sensitive to the mass of the SFDM, and reaching at the optimistic conclusion that eventually future experiments may be sensitive to $10^{-18}\rm{eV}.$ The most noticeable effect is a suppression of such features for lighter masses due to the  SFDM Jeans scale $k_J$.
As mentioned above, the suppression of the matter power spectrum for VFDM is characterized  by the same Jeans scale $k_J$ as for SFDM. Therefore,  we   expect a similar reduction of the VAO amplitude for a VFDM model for the corresponding masses.  Notice however the anisotropies  that are present for VFDM on short scales make the analysis for VFDM more complicated than for the SFDM case, and the anisotropies may  have an impact on the precise prediction of this effect.

\section{Conclusions} \label{sec: conclusions}
In this work we implemented  a  formalism  to compute the linear evolution of scalar cosmological perturbations with VFDM using the modified version of  the CLASS Boltzmann code. We analyzed the numerical results for the background and the scalar perturbation sector. At background level, we found an estimate  of a  lower bound to the vector field mass $m$ due to BBN constrains. 
At perturbation level, we focus on   the evolution of the cosmological perturbations  and on the shape of the matter power spectrum. In addition to presenting the numerical results, we performed a detail analytical study  of the evolution of the VFDM perturbations in the different relevant  regimes and  scales,  which allows to understand the obtained numerical results, and in particular the shape of the matter power spectrums. 
We  compared the matter power spectrum in VFDM models with the standard ${\rm CDM}$  model and the well-studied SFDM models.  As in the SFDM case, the VFDM model has  suppressed matter power spectrums with respect to CDM. We obtained the characteristic scale for the suppression is also given by the so-called Jeans scale $k_J = a_{osc}\sqrt{m H_{osc}}$, both analytically and numerically. For large enough values of the field mass  our results show that  there is no significant differences on   the matter power spectrum of the two models on large scales, but there are  considerable differences on the amplitudes of the  power spectrums for $k\sim k_J$, which in the future could help to distinguish the models.   For  smaller values of the field mass we note that the  early time anisotropies characterized by the background metric shear (sourced by the background vector anisotropic stress) has an impact on the  power spectrums on all scales. We present the   details of the study of such impact in  Appendix \ref{sec:shear_impact}. The results confirm the conclusion reached in \cite{Chase:2023puj}  on that it is necessary to take into account the metric shear in order to get a sufficiently  good approximation for the predictions at perturbative level. In particular, we obtained that for $m < 10^{-24}\rm{eV}$ the error in not considering the shear is $> 10 \%$.

In order to complete the implementation of the VFDM model in CLASS, we leave as future  work  the calculation and implementation of the CMB power spectrums. Due to the angular dependence of the Fourier modes of the matter power spectrums, it is necessary to modify the current formulae in the code. 
It would be also worth to perform a detail study of the vector and tensor sectors. Unlike in  CDM or SFDM models,  in VFDM models the scalar, vector and tensor perturbations  mix due to the presence of the background vector field.  As for SFDM, isocurvatuve modes are generically expected for VFDM (see for instance \cite{Kitajima:2023fun}). Therefore, another interesting extension of the code would be to allow for  generalized initial conditions that include VFDM isocurvature perturbations. 

The modified code presented in this paper is an important and necessary step  to assess to what extend the characteristic VFDM anisotropies can be used to distinguish this from SFDM. Indeed, the predictions of the linear evolution of the cosmological perturbations represent a crucial input for estimating  cosmological constrains precisely.   More work is needed to move forward in the characterization of the observable properties of VFDM models in the non-linear regime. The outputs of this code  can be used as a base to compute the initial conditions for N-body simulations.

Finally, in the future we hope to extend the code to study mixed models (involving both SFDM and VFDM components), and as a more challenging project, to  study spin 2 ULDM models.

\begin{acknowledgments}  
We would like to thank  Nahuel Mirón Granese, Mustafa Amin and Matías Zaldarriaga for discussions. This work has been supported by CONICET and UBA. We acknowledge the use of the xAct - xPand package for Mathematica \cite{Pitrou:2013hga,Brizuela:2008ra}. 
\end{acknowledgments}

\appendix

\section{Fluid variables}\label{appendix_fluid_variables}

We assume that the background metric shear is small in comparison with the Hubble parameter, so we solve for the shear perturbatelly by neglecting it  on the r.h.s of Einstein equations. 

For the background, the fluid variables of the vector field in Bianchi I are
\begin{align}
    \rho_A &= \frac{1}{2 a^4} \left[ A_{i}^{\prime}A_{j}^{\prime} + m^2 a^2 A_{i}A_{j} \right]\,\gamma^{ij}\,, \label{rho_A_background}\\[4pt]
    P_A &= \frac{1}{6 a^4} \left[ A_{i}^{\prime}A_{j}^{\prime} - m^2 a^2 A_{i}A_{j} \right] \,\gamma^{ij}\,, \label{background_pressure_expression}\\[4pt]
    {\Sigma^i}_j &= \frac{1}{a^4} \bigg[ \frac{1}{3}A_{k}^{\prime} A_{l}^{\prime}\, \gamma^{kl} {\gamma^i}_j - A_{k}^{\prime} A_{j}^{\prime} \, \gamma^{ik}  \\
    &\qquad+ m^2 a^2 (A^i A_j - \frac{1}{3}A^k A_k {\gamma^i}_j)  \bigg]\,. \nonumber 
\end{align}

For the perturbations, the fluid variables implemented in the code are
\begin{subequations} 
\begin{align}
    \delta &\rho_A  = \frac{1}{a^4} \bigg[A^{\prime}_L  \left(\delta A^{\prime}_L - i\,k\,\delta A_0\right) +  A_{t_1}^{\prime} \delta A_{t_1}^{\prime} + 2\rho_T a^4 \,\eta   \nonumber\\
    & + m^2 a^2 (A_L \,\delta A_L + A_{t_1} \,\delta A_{t_1}) \bigg] - \rho_L( h + 4 \eta )\,,\label{delta_rho_synch} 
    \end{align}
\begin{align} 
    \delta &P_A = \frac{1}{3a^4} \bigg[A^{\prime}_L (\delta A^{\prime}_L - i\,k\,\delta A_0) + A_{t_1}^{\prime} \delta A_{t_1}^{\prime} + 6a^4P_T \, \eta \nonumber \\ 
    &- m^2 a^2 (A_L \,\delta A_L + A_{t_1} \,\delta A_{t_1}) \bigg]  - P_L  ( h + 4 \eta )\,, \label{delta_P_synch} 
\end{align}
\begin{align} 
    (\rho_A &+ P_A) \theta_A = \frac{k^2}{a^4} A^{\prime}_T  \delta A_{T_1}-i \frac{m^2 \, k}{a^2} \, A_L\, \delta A_0\,, 
\end{align}
\begin{align} 
    (\rho_A &+ P_A) \delta\Sigma_{A\parallel}  = \frac{4}{3 a^4} \bigg[A^{\prime}_L(\delta A^{\prime}_L - i\,k\,\delta A_0)  - 3 a^4 P_L (h + 4 \eta) \nonumber\\
    &+ m^2 a^2 (A_{t_1} \,\delta A_{t_1} - A_L \,\delta A_L) \bigg] - 4 P_T \, \eta\,,  \label{delta_shear_synch}
\end{align}
\end{subequations}
where we defined $\rho_L = \rho_A \big|_{A_t = 0}$, $\rho_T = \rho_A \big|_{A_L = 0}$, $P_L = P_A \big|_{A_t = 0}$ and $P_T = P_A \big|_{A_L = 0}$.

\section{Change of variables} \label{appendix:change_of_variables} 

We can obtain the     equations of motions of the new variables introduced in Eq. (\ref{change_of_variables_A_xi_alphha}) by imposing  respectively that the time derivative of $\delta A_L$ and $\delta A_T$,  as defined  in  Eq. (\ref{change_of_variables_A_xi_alphhaAL}) and (\ref{change_of_variables_A_xi_alphhaAT}), to be  equal to the definition of $\delta \dot{A}_L$ and $\delta \dot{A}_T$ given in Eq. (\ref{change_of_variables_A_xi_alphhaBL}) and  (\ref{change_of_variables_A_xi_alphhaBT}). We then 
rewrite the equations of motion in term of this new set of variables. In this way  we get the following system of coupled differential equations: 
\begin{align}
    \xi^{\prime}_L &= \sin(\xi_L) + y + 2 \kappa \left[1 - \cos(\xi_L) + \frac{4 \sin(\xi_L)}{4 \kappa + y}\right] \\[4pt]
    &- e^{-\alpha_L} \sin\left(\frac{\xi_L}{2}\right) \bigg[ 2 \kappa\, \sin\left(\frac{\theta}{2}\right) (h + 8 \eta) \nonumber\\[4pt]
    &+ \cos\left(\frac{\theta}{2}\right) \left(\frac{8 \kappa}{4\kappa + y} (h + 8 \eta) + h^{\prime} + 8 \eta^{\prime} \right) \bigg]\,,\nonumber\\[9pt]
    \alpha^{\prime}_L &= \cos(\theta) - \cos(\xi_L) - 2 \kappa \bigg[\frac{4(1+\cos(\xi_L))}{4\kappa+y}\\[4pt]
    &\sin(\xi_L) \bigg]e^{-\alpha_L/2} \cos\left(\frac{\xi_L}{2}\right)\bigg[2 \kappa\, \sin\left(\frac{\theta}{2}\right) (h + 8 \eta) \nonumber
    \\[4pt]
    &+ \cos\left(\frac{\theta}{2}\right)\left( \frac{8\kappa}{4\kappa + y} (h + 8 \eta) + h^{\prime} + \eta^{\prime} \right) \bigg]\,,\nonumber\\[9pt]
    \xi^{\prime}_T &= \sin(\xi_T) + 2 \kappa(1 - \cos(\xi_T))  + y \\[4pt]
    &- e^{-\alpha_T/2} \sin\left(\frac{\theta}{2}\right)\sin\left(\frac{\xi_T}{2}\right)(h^{\prime} + 4 \eta^{\prime})\,,\nonumber\\[9pt]
    \alpha_T^{\prime} &= \cos(\theta) - \cos(\xi_T) - 2 \kappa \sin(\xi_T)\\[4pt]
    &- e^{-\alpha_T/2} \cos\left(\frac{\xi_T}{2}\right)\cos\left(\frac{\theta}{2}\right) (h^{\prime} + 4 \eta^{\prime})\,, \nonumber
\end{align}
where we recall that $\kappa =  \frac{k^2}{2 m a \mathcal{H}}$ and $y = 2 m a/\mathcal{H}$.

The previous system of equations can be solved numerically. However, as it is shown in section \ref{sec:implementation in class}, the change of variables proposed in Eq. (\ref{second_change_of_variables}) simplifies the equations of motion substantially.

\section{Impact of the shear tensor on the power spectrum} \label{sec:shear_impact}
As   mentioned in the main text, the VFDM model with adiabatic initial conditions requires the anisotropic  Bianchi I background  metric characterized by a shear tensor (see Eq. \ref{eq_shear_abundance_evolution}). 
In this section we calculate the error in the matter power spectrums one would make if the  vector field  were assumed to be in a FLRW background. As argued in \cite{Chase:2023puj}, this is mainly given by   the error produced by not including the term containing the shear tensor on the l.h.s of Einstein $0i$ equation (Eq. \ref{einstein_0i_synch}).  

\begin{figure}[H]
    \centering 
    \includegraphics[width=0.49\textwidth]{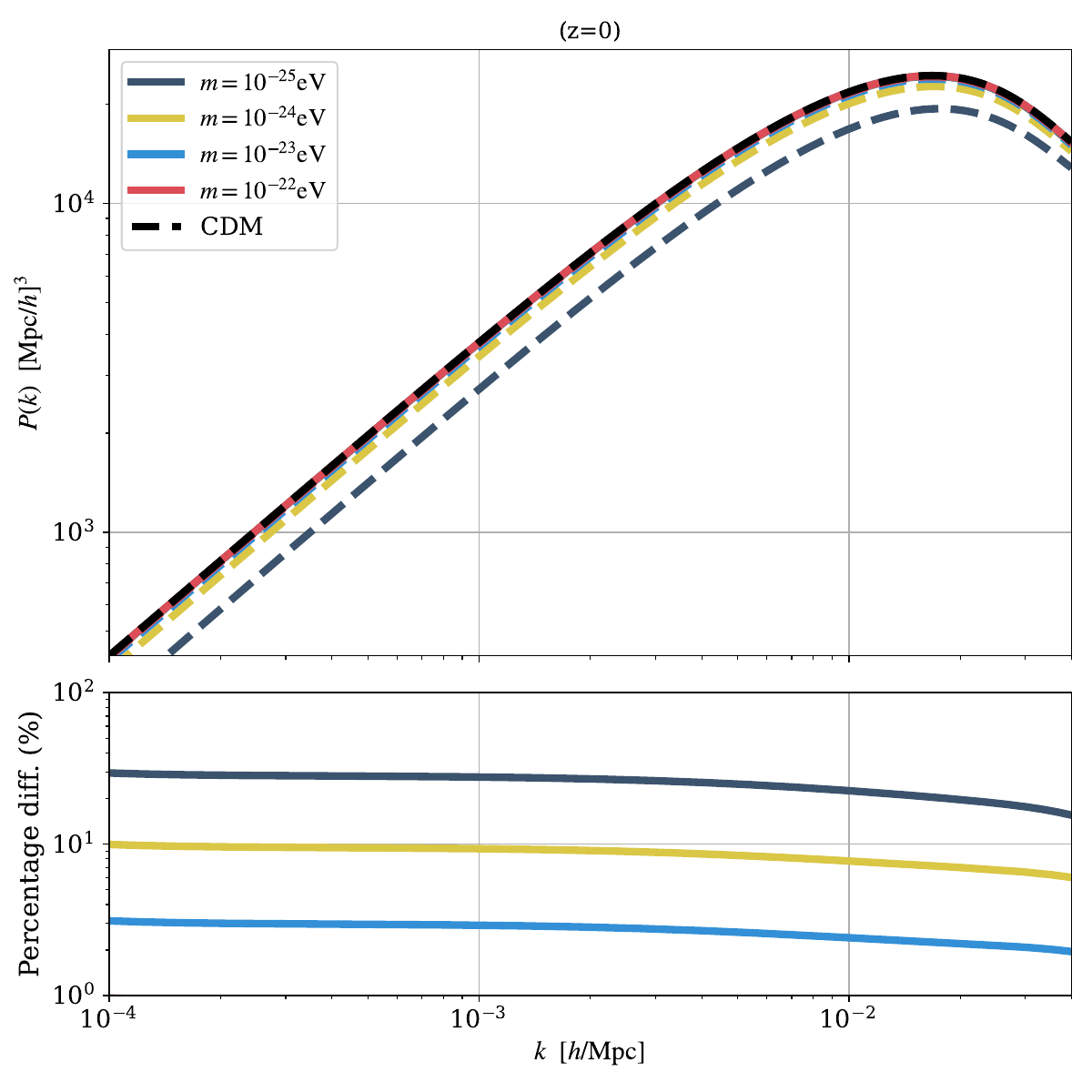}
    \caption{Matter power spectrum  considering the model in Bianchi I (solid lines) and in FRW (dashed lines), both calculated at $z=0$ and for purely longitudinal modes ($\gamma_k = 0$). In the lower panel we have the percentage difference between both models. We see that the error in not considering the shear is $> 1\%$ for masses $m < 10^{-22} \rm{eV}$ on large scales.}
    \label{fig: percentage_pk_shear_z=0}
\end{figure}

In figure \ref{fig: percentage_pk_shear_z=0} we show the matter power spectrum for $\gamma_k=0$ calculated in Bianchi I (that is, by taking into account the shear in Einstein $0i$ equation) in solid lines, and calculated in FLRW (dashed lines) for different masses of the vector field. In the lower panel we present the percentage error between the two cases. As it is expected, higher differences appear for the largest scales and smaller values for the masses. This can be understood by noting that at early times and outside the horizon the vector velocity gradient scales as $(k\tau)^2$, so it dominates over the radiation velocity gradient which scales as $k^4\tau^3$ in the same regime. This big contribution is exactly cancelled by the term containing the shear tensor on the l.h.s of Einstein $0i$ equation (see Eq. \ref{einstein_0i_synch}). Since the shear abundance increases for lower masses, we expect the effect to become more important  for the lowest  masses considered. In particular, we can see that for $m \sim 10^{-23}\rm{eV}$ the error in not considering the metric shear is $\sim 1 \%$, while for $m < 10^{-24}\rm{eV}$ the error is $> 10 \%$.

\section{Suppression in power spectrum and Jeans scale}\label{sec:jeans_scale} 

In this section we study the VFDM overdensity ($\delta_A$) during matter domination to understand the suppression in the mass power spectrum as a complement of Sec. \ref{sec:evolution_of_VFDM}.   For the masses considered in this calculation we have that $m \gg H$ in this regime, and the total equation of state is $w_T \sim w_A \sim 0$. As the field is highly oscillating the metric tensor averages to zero, so we neglect $\sigma_{ij}$. For non-relativistic scales we have that $y \gg \kappa$. In Fig. \ref{fig: percentage_pk_shear_z=0} we can see that VFDM behaves as CDM at large scales, while it has a suppression given at the denominated \textit{Jeans scale} $k_J$ which depends on the vector  mass. As we show next, this suppression is caused because $\delta_A$ oscillates on scales $k > k_J$, in contrast with its growing  behaviour at  scales $k < k_J$.

To understand the evolution of the power spectrum we start by calculating the effective fluid equations for the VFDM. This is done by deriving the corresponding fluid variables and using  the equation of motions. We then obtain 
\begin{subequations}
    \begin{align}
        \delta^{\prime}_A &= - \frac{\theta_A}{\mathcal{H}} - \frac{h^{\prime}}{2}\,, \\[4pt]
        \frac{\theta_A^{\prime}}{\mathcal{H}} &= - \frac{\theta_A}{\mathcal{H}} + \kappa^2 \delta_A - \kappa^2 (3 + \cos(2\gamma_k)) \eta \,.
    \end{align}
\end{subequations}
Using Einstein's temporal equation (Eq. \ref{einstein_temporal}), we can combine the two previous equations to obtain a second order   equation   for each of these fluid variables,
\begin{align}
    &\delta_A^{\prime\prime} + \frac{\delta_A^{\prime}}{2} + \left( \frac{k^4}{k_j^4} - \frac{3}{2}\right) \delta_A = 2\frac{k^2}{\mathcal{H}^2} \eta\,,
    \label{eq_mov_delta_A} \\[5pt]
    &\theta^{\prime\prime}_A + 3\, \theta^{\prime}_A + \left( \frac{k^4}{k_j^4} + 2\right) \theta_A = -\kappa^2 \frac{k^2}{\mathcal{H}}\eta\,.
    \label{eq_mov_theta_A}
\end{align}

The equation for the density contrast (Eq. \ref{eq_mov_delta_A}) has two different homogeneous solutions which are separated by the Jean's scale $k_J^2 = m a \mathcal{H}$. For $k \ll k_J$, $\delta_A$ has a growing mode, so we recover the $\rm{CDM}$ behaviour. For $k > \frac{3}{2} k_J$, $\delta_A$ has oscillatory solutions, so for these scales the perturbations do not grow and the power spectrum is suppressed. This can be seen in  Fig \ref{fig:delta_angles}, where in the top panel we have a growing mode ($k < \frac{3}{2} k_J$) and on the bottom panel we have a suppressed mode ($k > \frac{3}{2} k_J$). In the matter power spectrum this shows up as a characteristic suppression for  $k > \frac{3}{2} k_J$. On the other hand, the homogeneous equation for the velocity gradient (Eq. \ref{eq_mov_theta_A}) has in any case oscillating solutions.
We note that the vector Jeans scale does not depend on the Fourier mode orientation with respect to the vector field $\gamma_k$, in agreement with the results of Ref. \cite{Zhang:2023fhs}. 

To understand the particular solution of Eq. (\ref{eq_mov_delta_A}) we need the  equation of motion for the metric perturbation $\eta$. This equation can be obtained by deriving Einstein  temporal-spatial equation (Eq. \ref{einstein_0i_synch}) and using the fluid equations of the vector field, thus obtaining
\begin{equation}
    \eta^{\prime\prime} + \frac{3}{2} \eta^{\prime} = \frac{3}{2} \frac{k^2}{m^2 a^2} \left[ \delta - (3 + \cos(2\gamma_k))\eta\right]\,.
\end{equation}

For non-relativistic modes we have that $\eta \sim const$. Then, by inserting a constant solution for $\eta$ in Eq. (\ref{eq_mov_delta_A}), we can see that the particular solution grows logarithmically with the scale factor, so it is subdominant with respect to the homogeneous solution for $\delta_A$.

\bibliographystyle{ieeetr} 
\bibliography{references}

\end{document}